\documentclass[aps,amsmath,amssymb,10pt,twocolumn,footinbib,noshowkeys,showpacs,prd]{revtex4}

 \usepackage{epsfig}

\usepackage{amssymb}

\usepackage{amsfonts}
\usepackage{mathrsfs}
\usepackage{bm}
\usepackage{wick}
\usepackage{color}

\definecolor{Red}{rgb}{0.9,0.0,0.1}
\def\be{\begin{equation}}
\def\ee{\end{equation}}
\def\bea{\begin{eqnarray}}
\def\eea{\end{eqnarray}}
\def\bfl{\begin{flushleft}}
\def\efl{\end{flushleft}}
\def\bfr{\begin{flushright}}
\def\efr{\end{flushright}}
\def\bc{\begin{center}}
\def\ec{\end{center}}
\def\ben{\begin{enumerate}}
\def\een{\end{enumerate}}
\def\bit{\begin{itemize}}
\def\eit{\end{itemize}}

\def\dzn{,\kern-0.1em,}

\def\Lan{\langle}
\def\Ran{\rangle}

\def\Bz{\Big{[}}
\def\Bzz{\Big{]}}

\def\i{\text{i}}
\def\e{\text{e}}
\def\d{\text{d}}
\def\L{{\mathcal{L}}}
\def\H{{\mathcal{H}}}
\def\O{{\mathcal{O}}}

\def\Lan{\langle}
\def\Ran{\rangle}

\begin{document}

\title{Magnon Energy Renormalization and Low-Temperature Thermodynamics
of  O(3) Heisenberg Ferromagnets}

\author{Slobodan M. Rado\v sevi\' c }
\email{slobodan@df.uns.ac.rs}
%\author{
%Milica S. Rutonjski}
\author{ Milan R. Panti\' c}
\author{ Milica V. Pavkov-Hrvojevi\' c}
\author{Darko V. Kapor}

\affiliation{Department of Physics, Faculty of Sciences, University of Novi Sad, Trg Dositeja
 Obradovi\' ca 4, Novi Sad, Serbia}

\begin{abstract}

We present the perturbation theory for lattice magnon fields of
$D$-dimensional O(3) Heisenberg ferromagnet.
The effective Hamiltonian for the lattice magnon fields is obtained
starting from the effective Lagrangian,
with two dominant contributions that describe magnon-magnon interactions
 identified as a usual gradient term for the unit vector field
and a part originating in the Wess-Zumino-Witten term of 
effective Lagrangian. Feynman diagrams for lattice scalar fields with
derivative couplings are introduced, on basis of which we 
investigate the influence of magnon-magnon interactions
on magnon self-energy  and ferromagnet free energy.
We also comment appearance of spurious terms in the low-temperature
series for the free energy by examining magnon-magnon interactions and internal symmetry
of the effective Hamiltonian (Lagrangian).

\end{abstract}

%\keywords{O(3) Heisenberg ferromagnet, Effective Field Theory, 
%Lattice regularization}

\maketitle

%===============================================================================================================================================
\section{Introduction}
%===============================================================================================================================================

\label{intro}

Effective field  theory (EFT) is  well established
method for treating models exhibiting 
spontaneous symmetry breaking \cite{Burgess}  and
is applicable to low-energy part of
any system whose only massless excitations are Goldstone bosons \cite{AnnPhys}. 
Initially developed for the description of  low-energy sector of 
quantum chromodynamics (QCD),
where it is known by the name of the chiral perturbation theory
\cite{Weinberg,Gasser,Gerber,AnnPhys}, EFT was also adapted 
for the condensed matter problems
 \cite{PRD,Burgess,Brauner}.
In particular, an application of EFT to  
Heisenberg ferromagnet (HFM) has met with 
considerable success
\cite{RomanSoto,Hofmann1,Hofmann2,Hofmann3,Hofmann4,Hofmann5}. 
As it is well known, the ground state $|0\Ran$ of a HFM
 is determined by a preferred direction in the internal space, 
singled out by the total spin $\bm S = \sum_{\bm n} \bm S_{\bm n}$ 
and
small fluctuations
of the order parameter near the ground state are described by
 Goldstone bosons (magnons)
of spontaneously broken (spin) rotational symmetry.
In the low dimensional  ($D = 1,2$, $D$ being the dimensionality
of spatial lattice) 
isotropic  ferromagnets with short-range interactions,
 rotational symmetry of the Heisenberg Hamiltonian
is restored at finite temperatures \cite{Mermin} and 
spontaneous symmetry
breaking  is possible only if $D\geq 3$. (Henceforth we will always
assume $D\geq 3$ and nearest neighbor interaction.)

Spontaneous symmetry breaking (SSB) in HFM is distinguished from 
its Lorentz-invariant counterparts, since the number of 
 Goldstone particles is less than the number of broken generators.
In Lorentz-invariant theories, the number of Goldstone  bosons ($n_{\text{GB}}$),
as well as the number of Goldstone
 fields ($\pi^a, \;a = 1,2,\dots n_{\text{GB}}$), 
equals the number of broken symmetry generators ($n_{\text{BS}}$), i.e. 
 $n_{\text{GB}} = n_{\text{BS}} = \mbox{dim}(\mbox{G}) - \mbox{dim}(\mbox{H})$. Here G denotes
 spontaneously broken (internal) symmetry group of the underlying system 
and H is the symmetry group of the ground state.
Even though the symmetry breaking pattern in HFM 
is $\text O(3) \rightarrow \text O(2)$, the excitation spectrum
contains only one type of  magnon. 
This is related to the fact that ferromagnetic
magnons possess nonrelativistic dispersion $\omega \propto \bm k^2$
due to nonzero vacuum expectation values of  charge
 densities \cite{PRD,PhysLett,PRDnovi},
and a complex field $\psi \propto \pi^1 + \text i \pi^2$ describes a single particle 
\cite{Burgess,PRD,Hofmann1,Hofmann2,Hofmann3,Hofmann4,NielsChad}.
In other words, $\pi^1$ and $ \pi^2$ represent canonically conjugate variables
and not two distinct Goldstone fields \cite{Nambu}.
A general theorem on SSB in Lorentz-noninvariant systems \cite{PRDnovi} asserts
that twice the number $n_{\text{BS}} - n_{\text{GB}}$
equals  rank of the matrix $\rho$, defined by its elements
$\rho_{ij} = \lim_{V \rightarrow \infty} (-\text i/V) \Lan 0 |[ Q_i,Q_j ]|0 \Ran$,
where $V$ denotes the spatial volume of the system and $\{ Q_i \}$ is 
the set of broken generators (integrals of charge densities).
If, as usual,   the spontaneous magnetization aligns 
in the direction of positive $z$ axis, one finds
$\rho \propto \text{diag}[1,-1]$ and corresponding 
single ferromagnetic magnon.
This is in accordance with standard spin-wave theory.
The theorem was recently proved in \cite{JapanciPRL} using EFT
(see also related work in \cite{PhysLett,PRDnovi,Nambu,
JapanciPRL2,JapanciPRL3,Japanci4}) demonstrating
once again 
usefulness of the  effective Lagrangian method in the
theories without Lorentz invariance.

On the other hand, the thermodynamic properties of ferromagnets are usually
calculated by some variant of spin-wave theory.
The predictions of linear spin-wave theory (LSWT)
are reliable almost up to $T_{\text C}/2$ ($T_{\text C}$ denotes the Curie
temperature), but for 
quantitative description beyond
this temperatures one needs to incorporate the effects
of magnon-magnon interactions.
A  successful   theory of the spin-wave interactions in Heisenberg ferromagnets was
put forward by Dyson \cite{Dyson1,Dyson2}.
He had shown that the kinematical interaction, arising from the limitation
on the maximum number ($2S$) of the spin deviations on each lattice site, 
 may safely be ignored at temperatures not to close to $T_{\text C}$.
Dyson also demonstrated the weakness of dynamical magnon-magnon
interaction by calculating first order correction 
to the free energy and spontaneous magnetization
of 3D HFM, thus providing an  explanation for    the success  of LSWT.
The weakness of 
magnon-magnon interactions  reflects itself through the changes in the Bloch's
law. The first correction  due to magnon-magnon interactions is
 only of order $T^4$, compared to the leading term proportional
 to $T^{3/2}$.
Dyson's results were subsequently rederived using Holstein-Primakoff
bosons \cite{Oguchi} and the diagram technique for spin
operators \cite{VLP1,VLP2}. (See \cite{BKY,KaganovChubukov,Borovik,PSS} for a comprehensive  reviews
and list of original references.)
The discovery of high-temperature superconductivity (HTSC) revived interest
in the Heisenberg magnets. 
Since the mid of 1980-ties, a lot of work was
put in the understanding of  spin-wave interactions in systems
of localized spins. Theoretical constructions
 from this period, dealing explicitly with the
 Heisenberg ferromagnets, include the
modified spin wave theory (MSWT, see \cite{Takahasi1,Takahasi2,Takahasi3}), 
the large $N$ expansion of SU($N$) Heisenberg models and Schwinger boson mean 
field theory (SBMFT, see \cite{AuerbachArovas, SJKM, Chubukov}),
 the self-consistent spin wave theory \cite{Irkin}
 and renormalization group (RG) methods \cite{Kopietz,Irkin2}.
As in the earlier works  \cite{Oguchi,VLP1,VLP2,MBloch}, the 
authors of \cite{Takahasi1,Takahasi2,Takahasi3,AuerbachArovas,SJKM,Chubukov,Irkin}
had shown that a realistic description of the 
low temperature phase of HFM can be reached using
bosonic (or combined bosonic-fermionic \cite{Irkin}) 
representations of the spin operators within quartic approximation, or with the help of
appropriate mean field/random phase   approximations (MFA/RPA),
 without complicated mathematical
constructions of Dyson.
As an alternative to  boson/fermion 
Hamiltonians, obtained from one of many  representation of 
spin operators \cite{Garb}, several authors
developed the method of double time temperature Green's
functions (TGF). (A recent review and original references
can be found in \cite{Kunc}.) It is based on the equations
of motion for spin operators, which are turned
into a solvable system of algebraic equations by 
suitable linearization.
Known as the
 decoupling schemes 
in the language of the double time TGFs,
the linearizations  incorporate effects of
magnon-magnon interactions without any direct
reference to the nonlinear boson/fermion Hamiltonian, i.e. to
the magnon-magnon interaction operator.
One of the most frequently used approximations of this kind
is the one by Tyablikov (TRPA, see \cite{Kunc}), usually described
as the one in which correlations between $S^z$ and
$S^{\pm}$ operators from adjacent sites are neglected.
Magnon energy renormalization, a consequence of 
Tyablikov's approximation,  affects the low-temperature regime
of the theory. 
The low-temperature expansion of ferromagnetic  order parameter
for 3D lattice calculated in TRPA contains
so called spurious term $\propto T^3$, in disagreement with rigorous
results of Dyson. Despite this, 
TRPA yields reliable predictions in accordance with 
Mermin-Wagner theorem and closed system of equations for correlation
functions is often  tractable within %(using)
 standard numerical tools. Also, unique solution for critical
temperature with self-consistently determined parameters
 agrees  well with Monte Carlo simulations 
and experimental values \cite{Kunc}.  This aspect of true self-consistency
comes to be  important when real compounds are modeled 
by Heisenberg ferromagnet/antiferromagnet (see e.g.  \cite{PRB,PRB2,SSCCV,IJMPB,EPJB,SSCTNRPA}
for an application of TRPA to 
cuprates, iron pnictides
and manganites).

The standard theories of non-linear spin waves
mentioned in the previous paragraph are based on 
boson/fermion representations of spin operators.
Since the commutation relations for spin operators
and the dynamics of the spin system are fully 
satisfied only with exact boson/fermion Hamiltonian 
and corresponding Hilbert space,
theories of non-linear spin waves are
 sensitive to any form of approximation.
These include, e.g,  various mean-field approximations in the 
Heisenberg spin Hamiltonian
\cite{VLP1,VLP2,AuerbachArovas,SJKM}, or approximate
boson/fermion expressions for spin operators
\cite{Oguchi,Irkin}.
Similar remark holds for the theories based on the 
equations of motion for the spin operators
where the approximations are made in the commutator
for $S^\pm$ operators (see \cite{Englert,Stinchcombe}
and the section \ref{PD}).
All these simplifications basically alter the spin nature 
of $(S^\pm, S^z)$ operators in a manner that may not be obvious
within a given framework.
These problems do not arise in the EFT approach, since one works
with true magnon operators from the beginning.
All simplifications are directly related to
the piece of Lagrangian (Hamiltonian) describing magnon-magnon interactions.
This makes the influence of approximation
more transparent.

To the best of our knowledge, the perturbation theory
with lattice regularization
has not yet been applied to  the
EFT of a  ferromagnet. 
There are several reasons for using a lattice 
within Hamiltonian formalism.
First, 
unlike dimensional regularization,
frequently used within EFT framework \cite{AnnPhys,Hofmann2,Hofmann3,Hofmann4,Hofmann5}, 
the lattice regularization preserves   full discrete symmetry of the original 
Heisenberg Hamiltonian and it seems to be an appropriate method
to deal with system initially defined on a lattice. 
Second, we will address to some issues inaccessible  to the
continuum field theoretical methods of \cite{Hofmann1,Hofmann2,Hofmann3,Hofmann4,Hofmann5,
PRDnovi,PhysLett,NielsChad,Nambu,JapanciPRL,JapanciPRL2,JapanciPRL3,Japanci4}, such as the
influence of interactions on magnon energy renormalization
over the entire Brilouin zone. Further, it is the structure of interacting Hamiltonian
 for ferromagnetic magnons,
rather than the general form of  interacting Lagrangian, that reveals
certain simplifications in the diagrammatic calculation of the  magnon self-energy
and free energy of O(3) HFM.
Although it lacks some of the 
systematization capabilities of  continuum field theoretical approach, the
lattice regularized theory can provide us with a  useful
information not just about spin systems
but also on other standard techniques.
For example, by examining magnon mass renormalization in 
sections \ref{SE2Loop} and  \ref{RotSymLagr}, we reach a clear explanation
for spurious $T^3$ term in Tyablikov RPA.

The section \ref{PD} contains brief discussion on LSWT, the magnon mass renormalization
and its influence on the spontaneous magnetization. Some notation on the lattice theory,
such as the lattice Laplacian, are likewise introduced there. The effective  interaction
Hamiltonian of lattice magnons is derived in  the Section \ref{EffInter},
starting from the effective Lagrangian. 
The Feynman diagrams with colored propagators
and vertices, suitable for theories of lattice scalar fields with derivative couplings
are also defined in  the Section \ref{EffInter}.
Two-loop perturbation theory for lattice magnon self-energy
is presented in the Section \ref{SE2Loop}, while three-loop
analysis of the free energy is given in the Section \ref{3LFE}.
Results of Section \ref{RotSymLagr}, based on 
continuum field theoretic calculation supplement and 
 clarify findings of two preceding Sections.
An important feature of the effective Hamiltonian
is identification of the two types of magnon-magnon interaction different in origin.
The  careful discussion  in sections \ref{EffInter}--\ref{RotSymLagr} offers a
 new  answer for appearance of the spurious terms
in the low-temperature series and demonstrates the influence
of spin-rotation symmetry on the thermodynamic properties of O(3) HFM. 
Finally, some calculation details 
and an alternative formulation of the O(2) model 
of the Subsection \ref{EFTRPA} are collected in the Appendices.

%===============================================================================================================================================
\section{Preliminary discussion}\label{PD}
%===============================================================================================================================================

In this section we are  motivating approach
to be discussed in detail latter. 
Also, for the clarity of presentation, we find it convenient
to introduce some notation on lattice fields
before general perturbation theory.

First, it is instructive to rewrite the
Hamiltonian of Heisenberg ferromagnet
with nearest neighbor interaction $J$ on a 
$D$ dimensional lattice
%-------------------------------------------------------------------------------------------------------------------
\bea
H & = & - \frac{J}{2}\sum_{\bm{n},\; \bm \lambda} 
  \bm S_{\bm n} \cdot \bm S_{\bm n + \bm \lambda},  
 \label{HFM1}
\eea
%-------------------------------------------------------------------------------------------------------------------
in terms of the discrete   Laplacian
%------------------------------------------------------------------------------------------------------------------------
\bea
\hspace*{-0.5cm}H & = & 
 -\frac{1}{2}  \;\frac{J Z_1 |\bm \lambda|^2 }{2D} 
 \sum_{\bm x}  \bm S_{\bm x}   \cdot 
\nabla ^2 \bm S_{\bm x} 
 -  \frac{J S (S+1)Z_1 N}{2}. \label{HFM2}
\eea
%------------------------------------------------------------------------------------------------------------------------
Here $[ S_{\bm x}^{i},  S_{\bm y}^{j} ] = 
\i \epsilon_{i j k}
S_{\bm x}^{k}\Delta(\bm x - \bm y) $, $\bm S^2 = S (S+1)$, the
lattice Laplacian is  (See e.g. \cite{LatLapl})
%------------------------------------------------------------------------------------------------------------------------
\bea
\hspace*{-0.4cm}\nabla^2 \phi(\bm x)   = \frac{2D}{Z_1 |\bm \lambda|^2} 
 \sum_{   \bm \lambda   } \Big{[}\phi(\bm x + \bm \lambda) 
-  \phi(\bm x)\Big{]},
\label{OpIzLapA}
\eea
%------------------------------------------------------------------------------------------------------------------------
$\{ \bm \lambda \}$ are the vectors that connect given site $\bm x$
with its $Z_1$ nearest neighbors and $N$ is the total number of lattice sites.
As we are mainly interested in finite temperatures, 
imaginary time formalism is used throughout the paper
(unless otherwise stated).
Employing $-\partial_{\tau}S_{j}(\bm n,\tau) = [S_{j}(\bm n,\tau),H],
j = 1,2,3$,
equation of motion for $\bm S(\bm x,\tau)$ is found to be
(imaginary time arguments are suppressed)
%------------------------------------------------------------------------------------------------------------------------
\bea
-\partial_\tau \bm  S(\bm x) &  = & - \frac{\i}{2} \; \frac{J Z_1 |\bm \lambda|^2 }{2D}
\Bz \Big{(} \nabla^2 \bm S(\bm x)  \Big{)} \times \bm S(\bm x) \nonumber \\
& - &
\bm S(\bm x) \times  \nabla^2 \bm S(\bm x)    \Bzz  \label{LLEq} .
\eea
%------------------------------------------------------------------------------------------------------------------------
Eq. (\ref{LLEq}) is just the  lattice version of imaginary time 
Landau-Lifshitz equation for operators $\bm S(\bm x, \tau)$. 
It can be solved in a linear approximation. 
Assuming the long range order (LRO), we may set $S^z(\bm n) \approx S$
~\footnote{As stated in the Introduction,  approximations 
of this kind are expected to be valid only for $D\geq 3$ if
$T \neq 0$.}. 
In this approximation, equation of motion for $S^+(\bm x) = S^x(\bm x) + \i S^y(\bm x)$
takes the form of the imaginary time equation for Schr\"odinger field on the lattice
%-------------------------------------------------------------------------------------------------------------------
\bea
-\partial_\tau   S^+(\bm x) &  = & - \frac{1}{2 m_{\text{LSW}}}
 \nabla^2  S^+(\bm x)       \label{SEq1},
\eea
%-------------------------------------------------------------------------------------------
where we have defined
%-------------------------------------------------------------------------------------------------------------------
\bea
 m_{\text{LSW}} = \frac{2D}{2 J S Z_1 |\bm \lambda|^2}    \label{MLSW}.
\eea
%-------------------------------------------------------------------------------------------------------------------
Similar equation holds for $S^-(\bm x) = S^x(\bm x) - \i S^y(\bm x)$.
Simultaneously, the linearized commutation relations for $S^\pm$ operators   read
%-------------------------------------------------------------------------------------------------------------------
\bea
\left[ \frac{S^+(\bm x)}{\sqrt{2 S}}, 
\frac{S^-(\bm y)}{\sqrt{2 S}}  \right] = \Delta (\bm x - \bm y) \label{KR1}.
\eea
%-------------------------------------------------------------------------------------------------------------------
Comparing (\ref{KR1}) to the usual form of equal-time commutation relations,
$[\psi(\bm x), \psi^\dagger(\bm y)]
= v_0^{-1} \Delta(\bm x - \bm y)$, where $v_0$ denotes
volume of the primitive cell, we see that in this approximation the Heisenberg 
ferromagnet is described by the bosonic lattice Schr\"odinger fields
%-------------------------------------------------------------------------------------------------------------------
\bea
\psi(\bm x,\tau) = \frac{S^+(\bm x,\tau)}{\sqrt{2 S  v_0}},\;\;\;\;
\psi^\dagger(\bm x,\tau) = \frac{S^-(\bm x,\tau)}{\sqrt{2 S  v_0}},
\label{PsiDef1}
\eea
%-------------------------------------------------------------------------------------------------------------------
which annihilate and create magnons at lattice site $\bm x$, respectively.
Schr\"odinger field interpretation can be further justified by 
solving equation (\ref{SEq1}) and constructing a diagonal Hamiltonian.
Finding plane-wave solutions \cite{Fetter} of (\ref{SEq1}) 
%-------------------------------------------------------------------------------------------------------------------
\bea
\hspace*{-0.4cm}\psi(\bm x,\tau) = 
\int_{\bm k} a_{\bm k}\; \e^{\i \bm k \cdot \bm x - \omega(\bm k) \tau},
\;\;\; \int_{\bm k} \equiv  \int_{\text{IBZ}} \frac{\d^D \bm k}{(2 \pi)^D}
\label{RPAPW}
\eea
%-------------------------------------------------------------------------------------------------------------------
and using eigenvalues of the lattice Laplacian
%-------------------------------------------------------------------------------------------------------------------
\bea
&&\nabla^2 \exp [\i \bm k \cdot \bm x]  =  
-\frac{2 D}{|\bm \lambda|^2} [1-\gamma_{D}(\bm k)] \exp [\i \bm k \cdot \bm x] 
\label{kSq} \\
& \equiv &  - \widehat{\bm k}^2 \exp [\i \bm k \cdot \bm x], \;\;\;\;\;\;
 \gamma_{D}(\bm k) = Z_1^{-1} \sum_{\{ \bm \lambda \}} \exp [\i \bm k \cdot \bm \lambda],
 \nonumber
\eea
%-------------------------------------------------------------------------------------------------------------------
we find the magnon dispersion 
%-------------------------------------------------------------------------------------------------------------------
\bea
\omega_{\text{LSW}}(\bm k) & = & \frac{\widehat{\bm k}^2}{2 m_{\text{LSW}}} = 
 J Z_1 S [1 - \gamma_{D}(\bm k)] \label {OmegaLSW},
 \eea
%-------------------------------------------------------------------------------------------------------------------
and diagonal magnon Hamiltonian in the linear approximation
%-------------------------------------------------------------------------------------------------------------------
\bea
H_0  & = &  - \frac{v_0}{2 m_{\text{LSW}}} 
\sum_{\bm x}  \psi^\dagger(\bm x)  \nabla^2 \psi(\bm x) 
- E_0 \label{DiagLSW} \\
& = & V \int_{\bm k} \omega_{\text{LSW}}(\bm k)\; n_{\bm k}
- E_0,\;\;\;  E_0 = \frac{J Z_1 N S^2}{2} \nonumber,
\eea
%-------------------------------------------------------------------------------------------------------------------
where $V n_{\bm k} = a_{\bm k}^\dagger a_{\bm k}$ and
$V = (2 \pi)^D \delta(\bm k - \bm k) = N v_0$.
$a_{\bm k}$ and $a_{\bm k}^\dagger$ are standard bosonic operators obeying
 commutation relations $[a_{\bm k}, a_{\bm q}^\dagger] = 
(2 \pi)^D \delta(\bm k - \bm q)$.
Operating on the vacuum  $|0\Ran$, $a_{\bm p}^\dagger$ creates one-magnon  state
$a_{\bm p}^\dagger |0\Ran = |\bm p\Ran$. These states are   normalized as 
$\Lan \bm p | \bm q \Ran = (2 \pi)^D \delta(\bm p - \bm q)$ \cite{PRD}.
We may now identify $m_\text{LSW}$  as (bare) mass of the lattice field quanta i. e. magnons.
We shall continue to refer to $m_\text{LSW}$ as a magnon mass
because of the nonrelativistic form of the dispersion relation (\ref{OmegaLSW}),
even though ferromagnetic magnons are "massless" from the point of view of the 
Goldstone theorem.

The diagonal Hamiltonian makes  thermodynamical properties
of a ferromagnet trivial to calculate.
For example, at low temperatures, the spontaneous magnetization per
lattice site
is found to be a vacuum expectation value of
%-------------------------------------------------------------------------------------------------------------------
\bea
S_{\bm x}^z & = & \sqrt{S(S+1)- [S^x_{\bm x}]^2-[S^y_{\bm x}]^2}
 \approx S - \frac{S_{\bm x}^- S_{\bm x}^+}{2S} 
\nonumber \\
& = & S - v_0 \psi^\dagger(\bm x) \psi (\bm x).
\eea
%-------------------------------------------------------------------------------------------------------------------
Written in terms of the thermal propagator for Schr\"{o}dinger field
%-------------------------------------------------------------------------------------------------------------------
\bea
\hspace*{-0.2cm}D(\bm x - \bm y, \tau_x - \tau_y) & = & 
\Lan \mbox{T} \left\{ \psi(\bm x,\tau_x) \psi^\dagger (\bm y,\tau_y) 
 \right\} \Ran_0  \label{Prop} \\
& = &  \frac 1\beta \sum_{n = -\infty}^\infty 
\int_{\bm q} \frac{\e^{\i \bm q \cdot (\bm x - \bm y) - \i
 \omega_n (\tau_x-\tau_y)}}{\omega(\bm k) - \i \omega_n}, \nonumber
\eea
%-------------------------------------------------------------------------------------------------------------------
it is
%-------------------------------------------------------------------------------------------------------------------
\bea
\Lan S^z \Ran  = S- v_0 \int_{\bm q} \Lan n_{\bm q} \Ran_0 
= S - v_0 D(0).\label{SzPropagator}
\eea
%-------------------------------------------------------------------------------------------------------------------
Here $D(0)$ denotes the propagator evaluated at the origin,
$\Lan n_{\bm q} \Ran_0$ is the free-magnon Bose distribution
and we have used the sum rule \cite{Fetter}
$\beta^{-1} \sum_n [\omega_0({\bm p})-\i \omega_n]^{-1} = \Lan n_{\bm p}\Ran_0$.
Results (\ref{MLSW})-(\ref{DiagLSW}),
which define the lattice theory of  free Schr\"{o}dinger field, 
are easily seen to be those 
of standard linear spin waves (LSW).

\vspace*{0.5cm}

The question of how to incorporate the effects of magnon-magnon interactions
into equations like (\ref{SzPropagator}) has long history
and long list of answers. They are grouped in several categories
as described in the Introduction.
The primary goal of the present paper  is to
show that thermodynamic properties of a $D$ dimensional
O(3) Heisenberg ferromagnet may be calculated
within formalism of interacting lattice Schr\"{o}dinger field,
based on the effective Lagrangian.
We will show, e.g., that the spontaneous magnetization of 
O(3) HFM to the first order in $1/S$,  can be written as
$S - v_0 G(0)$, where $G(0)$ is the magnon field Green's
function calculated to the one loop.
Also, in Sections  \ref{EffInter}--\ref{RotSymLagr}
we develop the perturbation theory capable for calculating
both micro and macro properties of O(3) HFM.

%\vspace*{0.5cm}

Returning to the magnon dispersion, it is easily seen that approximation 
$S^z(\bm n) \approx \Lan S^z \Ran$ in (\ref{LLEq})
eventually leads to TRPA result with the magnon energies
%-------------------------------------------------------------------------------------------------------------------
\bea
\hspace*{-0.9cm}\omega_{\text{RPA}}(\bm k) & = & \frac{\widehat{\bm k}^2}{2 m_{\text{RPA}}},\;\;\;\;
  m_{\text{RPA}} = \frac{2D}{2 J \Lan S^z \Ran Z_1 |\bm \lambda|^2} 
 \label {OmegaTRPA}.
 \eea
%-------------------------------------------------------------------------------------------------------------------
As happens in LSWT, the final result in TRPA
contains no information about the short range 
fluctuations (SRF) of the order parameter if the operator
$S^z(\bm n)$ is replaced with the site independent average $\Lan S^z \Ran$.
(A discussion about the role of SRF can be found in a recent review \cite{Plakida}.)
In spite of that, TRPA  incorporates certain type
of magnon-magnon interaction that renormalizes magnon mass
according to (\ref{OmegaTRPA}). Since the approximation is made
directly in the equation of motion, an explicit form
of magnon-magnon interactions that yield (\ref{OmegaTRPA}) can't be 
deduced in TGF formalism.  
Tyablikov's result (\ref{OmegaTRPA}) for HFM was subsequently  re-derived by
linearizing the commutation relations for Fourier components of $S^{\pm}_{\bm n}$ 
operators \cite{Englert,Stinchcombe} similarly as  in
this section, using the
perturbation theory for self-consistent mean field approximation
\cite{VLP1,VLP2},
various diagram techniques
for spin operators 
 \cite{BKY,Fishman},
drone-fermion for $S = 1/2$ and $S = 1$
\cite{Spencer} and pseudofermion representation for
spin $S=1/2$ ferromagnets \cite{KineziRPA}. 
Although both the spin-operator diagram technique and the pseudo/drone fermion 
representations eventually yield TRPA result (or improve it), non of 
these approaches 
 describes HFM as a system of interacting magnons
built on LSWT as the non-interacting  theory, i.e.
using the perturbation theory for interacting magnon fields
without additional MFA/RPA approximations.
The issue of magnon-magnon interactions in TRPA
can be resolved by interpreting TRPA as a certain type of EFT. As a corollary, we will give 
a clear answer for the spurious $T^3$ term
of Tyablikov.

%\vspace*{2cm}
%===============================================================================================================================================
\section{Effective interaction of the lattice magnon fields}\label{EffInter}
%===============================================================================================================================================

We have seen in the previous section that the LSWT description
of HFM is equivalent to a theory of  free lattice Schr\"{o}dinger field.
The rest of the present paper will be devoted to the 
influence of magnon-magnon interactions on microscopic and macroscopic
properties of a ferromagnet. The simplest choice of the interaction  
for the Schr\"odinger field, with the Hamiltonian
density $\propto [\psi^\dagger (\bm x) \psi (\bm x)]^2$ simply won't work because the vertices
of $[\psi^\dagger (\bm x) \psi (\bm x)]^2$ interaction carry no momentum, so it can not 
renormalize the mass of a ferromagnetic magnon.
The correct form of the effective interaction
is  most easily formulated in terms of the Goldstone  fields
$\pi^a(\bm x)$.

\subsection{Effective Lagrangian}

As noted in the Introduction, the general effective Lagrangian
is written in terms of  Goldstone
fields $\pi^a (x), a = 1,2 \dots \text{dim(G)} - \text{dim(H)}$.
Various terms appearing in the effective Lagrangian are
organized in the  powers of   momenta of the Goldstone fields.
The leading order Lagrangian collects all contributions of the order $\bm p^2$.
If the system is invariant under parity, which is the case with the
Heisenberg Hamiltonian (\ref{HFM1}), only terms with even powers
of momenta are permitted. The next-to-leading order Lagrangian then
contains all contributions of the order $\bm p^4$ and so on.
Translated into the direct space, the powers of momenta correspond to the
derivatives. The effective Lagrangian
is constructed by adding terms with increasing number of
derivatives of the Goldstone fields, with the lowest order term
containing two derivatives. 
It should be noted that for systems whose
massless excitations  characterize
nonrelativistic dispersion $\omega \propto \bm p^2$, such is HFM, 
single time derivative 
counts as $\bm p^2$, i.e. as two spatial derivatives or a
single power of temperature \cite{PRD,Hofmann2,Hofmann3}.
Expansion in the powers of
momentum is   always terminated at some finite order
and, beside Goldstone fields and their derivatives, the
effective Lagrangian
also includes several coupling constants whose values are not specified 
by the symmetry requirements. They can be determined by comparison
of predictions of EFT with numerical simulations, experimental results
or by matching with detailed microscopic calculations \cite{Burgess,AnnPhys}. 
When the effective Lagrangian is constructed, a straightforward application
of Feynmann rules enables one to calculate the correlation functions,
partition function etc.
For a Heisenberg ferromagnet $\text G = \text O(3), \text H = \text O(2)$, and the spontaneous
symmetry breaking is accompanied by two real Goldstone fields $\pi^1(x)$ and $\pi^2(x)$.
However, the ferromagnetic magnons possess nonrelativistic dispersion
relation ($\omega \propto \bm k^2$) and a complex
field $\psi \propto \pi^1+ \i \pi^2$ describes a single magnon
 \cite{PRD,Hofmann1,Hofmann2,Hofmann3,Burgess,NielsChad,JapanciPRL,Nambu}. 

Effective Lagrangian for HFM was introduced in \cite{PRD,RomanSoto,Wiese}
(see also earlier works \cite{Jevicki,Klauder,WenZee}),
and a detailed derivation of the partition function up to the 
three loops using continuum approximation and the dimensional regularization, 
resulting with the leading corrections to  Dyson's analysis of 3D HFM,
can be found in \cite{Hofmann3} (see also \cite{Hofmann1,Hofmann2} and
\cite{Hofmann4,Hofmann5} for corresponding analysis of the two dimensional ferromagnet).
In the present paper, a slightly modified path will be followed.
As one of our interests lies in the mass renormalization of the lattice magnons,
we wish to preserve the full discrete symmetry of  lattice spin
Hamiltonian (\ref{HFM1}).
Because of that, we find it more convenient to work in the Hamiltonian
formulation of lattice field theory \cite{Hamer,Kogut}, leaving 
only (imaginary) time coordinate continuous.  Therefore, the first task is
to construct the effective Hamiltonian that describes interactions of
lattice magons with nonrelativistic dispersion.
Details for lattice regularization of the Lorentz-invariant effective field theory 
 can be found e.g. in  \cite{Smilga,Lewis}.

The leading order real time effective Lagrangian of O(3) ferromagnet is \cite{PRD,Wiese,Jevicki}
%-------------------------------------------------------------------------------------------------------------------
\bea
\L_{\text{eff}} = \Sigma \frac{\partial_t U^1 U^2 - \partial_t U^2 U^1}{1+U^3}
-\frac{F^2}{2} \partial_{\alpha} U^i \partial_{\alpha} U^i, \label{EffLagr}
\eea
%-------------------------------------------------------------------------------------------------------------------
where two magnon fields are collected into the unit vector $U^i:=[U^1,U^2,U^3]^{\text T}
\equiv [\bm \pi (x),U^3(x)]^{\text T}$, $\Sigma = N S/V$ is the spontaneous
magnetization per unit volume at $T = 0$K
and $F$ is a constant. The first part of Lagrangian is
usually denoted as Wess-Zumino-Witten (WZW) term. It gives rise to the
Berry phase \cite{Wen} and is responsible for the
classical dispersion of the ferromagnetic magnons ($\omega \propto \bm k^2$).
The presence of WZW term makes Lagrangian rotationally
invariant only up to the total derivative. As it will be shown,
the inclusion of  magnon-magnon interactions
arising from WZW term is crucial for correct
low-temperature description of O(3) HFM.
The next-to-leading order Lagrangian contains 
terms such as $l_1 (\partial_\alpha U^i \partial_\alpha U^i)^2$, 
$l_1 (\partial_\alpha U^i \partial_\beta U^i)^2$, 
$l_3 U^i \Delta^2 U^i$, or  $l_4 \partial^2_\alpha U^i \partial^2_\alpha U^i$  
with  arbitrary coupling constants $l_1, \dots l_4$ \cite{Hofmann2,Hofmann3}.
These $\O(\bm p^4)$ terms shall not be directly included in the effective Hamiltonian.
Instead, higher order momentum contributions will appear naturally
in a lattice regularized theory.
This regularization, however,  
restricts possible choices for higher order terms (See section \ref{RotSymLagr}).

\subsection{Transition to interaction picture}

The use of perturbation theory requires clear separation
between the 
free-magnon part, which must be identified with (\ref{DiagLSW}), and the interaction  part 
of the Hamiltonian \cite{WeinbergQTF1}. 
To extract them from the Lagrangian (\ref{EffLagr}),
we may rewrite it
in terms of the complex field $\psi = \sqrt{\Sigma/2} [\pi^1 + \i \pi^2]$
which describes the physical magnon,
and follow the standard canonical prescription. However, this is not 
the most efficient way to construct interaction picture.
The WZW term modifies canonical momentum,
from $\i \psi^\dagger$ of noninteracting theory, to $ 2\i \psi^\dagger /[1+U^3]$, 
where $U^3 = \sqrt{1-(2/\Sigma)\psi^\dagger \psi}$.
Consequently, the complex fields $\psi$ and $\psi^\dagger$ that enter Hamiltonian
are not those obeying equal-time commutation relations.
As the connection between canonical momentum and $\psi^\dagger$
is highly nonlinear it can be solved for $\psi^\dagger$ only iteratively.
Because of that, an important part of magnon-magnon interactions
is not manifest in the Hamiltonian, since it enters the quantum theory
 through the failure of $\psi$ and $\psi^\dagger$ 
to satisfy canonical Schr\"{o}dinger-field commutation
relations. This is reminiscent of the situation dealt
with in the spin-operator approach to Heisenberg magnets:
the commutation relations governing the dynamics of
system are neither Bose nor Fermi type
and the  interaction  is generated by
expanding localized spins operators in terms of boson/fermion operators
\cite{Oguchi,BKY,Garb,Irkin,Borovik}.
For the present purposes, however, it is desirable to have an explicit form of
the magnon-magnon interaction.

A different strategy \cite{WeinbergQTF1} makes use of the equation of motion,
which in the present case is the Landau-Lifshitz equation \cite{PRD},
%-------------------------------------------------------------------------------------------------------------------
\bea
\partial_t U^a + \frac{F^2}{\Sigma} \; \varepsilon_{aij} (\Delta U^i) U^j = 0, \label{LLEqCont}
\eea
%-------------------------------------------------------------------------------------------------------------------
to eliminate $\dot \pi^1$ and $\dot \pi^2$ from the interaction
part of the  Lagrangian (\ref{EffLagr}). Here
$\Delta = \partial_\alpha \partial_\alpha$.
In this manner we find the free-magnon Lagrangian
%-------------------------------------------------------------------------------------------------------------------
\bea
\L_{\text{free}} = \frac{\Sigma}{2}  \left[ \partial_t \pi^1 \pi^2 - \partial_t \pi^2 \pi^1 \right]
+\frac{F^2}{2} \bm \pi \cdot \Delta \bm \pi, \label{Lfree}
\eea
%-------------------------------------------------------------------------------------------------------------------
and the interaction piece
%-------------------------------------------------------------------------------------------------------------------
\begin{widetext}
\bea
\L_{\text{int}} = \frac{F^2}{2} \left[ 2- \bm \pi^2 - \sqrt{1-\bm \pi^2}   \right]
\Delta \sqrt{1-\bm \pi^2} - 
\frac{F^2}{2} \frac{1-\sqrt{1-\bm \pi^2}}{1+\sqrt{1-\bm \pi^2}}  \sqrt{1-\bm \pi^2} \; \bm \pi \cdot \Delta 
\bm \pi \label{Lint}.
\eea
\end{widetext}
%-------------------------------------------------------------------------------------------------------------------
Canonical interacting quantum theory can now be easily
constructed starting from $\L_{\text{free}}$ and $\L_{\text{int}}$.
To perform two-loop calculations for the self-energy ant three-loop
calculations for the free energy, we need to retain tjhe magnon-magnon
interaction up to and including six magnon operators. By expanding
(\ref{Lint}) we find that terms with six $\pi^a$ operators
precisely cancel, in contrast to a Lorentz-invariant theory
\cite{Gerber,Smilga}. Remaining four-magnon terms are then collected 
to \footnote{Needless to say, the same form of   $\H_{\text{int}}$ is found
by expressing $\psi^\dagger$ in terms of $\psi$ and canonical momentum,
as outlined at the beginning of this subsection.}
%-------------------------------------------------------------------------------------------------------------------
\bea
\hspace*{-0.4cm}\H_{\text{int}} = \frac{F^2}{8} \bm \pi^2
 \left[
 \bm \pi \cdot \Delta \bm \pi - \Delta \bm \pi^2 \right].
  \label{IntEffCont}
\eea
%-------------------------------------------------------------------------------------------------------------------

Finally, by putting the free Hamiltonian  and interaction part
(\ref{IntEffCont}) on the lattice, we obtain the effective
Hamiltonian for  lattice magnon 
fields 
%-------------------------------------------------------------------------------------------------------------------
\bea
H_{\text{eff}} & = & H_0 + H_{\text{int}}, \label{HamEffLatt0} \\
H_0 &  = & -\frac{1}{2 m_0} v_0 \sum_{\bm x}
 \psi^\dagger(\bm x) \nabla^2 \psi(\bm x),\;\;\; m_0 = \frac{\Sigma}{2 F^2}        \nonumber 
\eea
\bea
H_{\text{int}} & = & \frac{F^2}{8}  v_0 
\sum_{\bm x} \bm \pi^2(\bm x) 
 \left[ \bm \pi(\bm x) \cdot \nabla^2 \bm \pi(\bm x) - \nabla^2 \bm \pi^2(\bm x)  \right]\nonumber \\
 & \equiv & H_{\text{int}}^{(\text I)} + H_{\text{int}}^{(\text{II})}  \label{HamEffLatt1},
\eea
%-------------------------------------------------------------------------------------------------------------------
where $\nabla^2$ denotes the lattice Laplacian and the lattice Schr\"odinger field is 
($\Sigma  = S/v_0$ in a lattice theory)
%-------------------------------------------------------------------------------------------------------------------
\bea
\psi(\bm x) = 
\sqrt{\frac{S}{2 v_0}}\left[\pi^1(\bm x) + \i \pi^2(\bm x)\right] .
\eea
%-------------------------------------------------------------------------------------------------------------------
%-------------------------------------------------------------------------------------------------------------------
\begin{figure}
\includegraphics[scale=1.0]{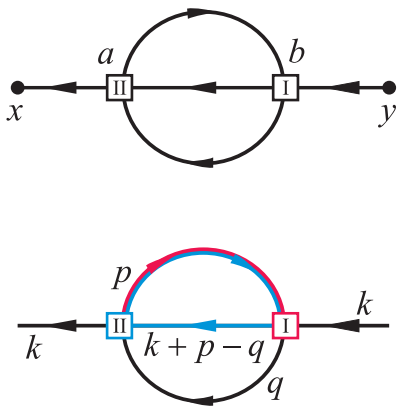}
\caption
{\label{fig3} (Color online) Coordinate space representation 
of a diagram contributing to the 
self-energy of lattice magnon field (top)
and one of its contractions in momentum space
(bottom). Vertices from $H_{\text{int}}^{(\text I)}$ and 
$H_{\text{int}}^{(\text{II})}$ are denoted by red
and blue squares, respectively.
The  blue line denotes propagator affected by
lattice Laplacians of $H_{\text{int}}^{(\text{II})}$. 
The double
colored line represents propagator acted on by Laplacians 
of both $H_{\text{int}}^{(\text I)}$ and 
$H_{\text{int}}^{(\text{II})}$.
}
\end{figure}
%-------------------------------------------------------------------------------------------------------------------
In what follows,  $\psi$ will always be written to 
the right in expressions like $\bm \pi \cdot \bm \pi$.
   $H_0$ is basically LSWT Hamiltonian (\ref{DiagLSW}) and
with this choice for $\L_{\text{free}}$ and $\L_{\text{int}}$ (i.e.
$H_0$ and $H_{\text{int}}$), $\psi$ and $\psi^\dagger$ do
satisfy canonical commutation relations for Schr\"{o}dinger field. 
Unlike in its continuous counterpart, the discrete symmetry of the original 
Hamiltonian (\ref{HFM1}) that modifies magnon dispersion
in higher orders of momentum is fully preserved in (\ref{HamEffLatt0}) and (\ref{HamEffLatt1}). Hence,
all higher order terms in momentum, i.e. in spatial derivatives,
that resolve the lattice structure at the same time  describing the free magnons are
collected in $H_0$. This fact simplifies further calculations.
Magnon-magnon interactions in accord with lattice structure and internal symmetries,
to the order considered here,
are collected in $H_{\text{int}}$.
One can choose constant $F^2$  to be 
$\Sigma/(2m_{\text{LSW}}) = J S^2 Z_1 |\bm \lambda|^2/(2 D v_0)$, so that
the energy of free lattice magnons is measured in units of $J$ 
\footnote{We note that, with $F^2=J S^2 Z_1 |\bm \lambda|^2/(2 D v_0)$,
interaction Hamiltonian is
 independent of spin magnitude $S$  
when written
in terms of magnon fields $\psi$.
In other words, the weakness of magnon-magnon interaction
$H_{\text{int}}$  is not controlled by $1/S$ expansion,
as opposed to what is often implicitly assumed in
calculations based on boson representations of spin
operators
(see, e.g. \cite{Oguchi, BKY,Borovik,Zitomirski}).
The Goldstone bosons are derivatively coupled and thus interact weakly at
low momenta.
One should not fail to notice that similar  arguments in the spirit
of EFT 
were also given by Dyson \cite{Dyson1}.}, as
in LSWT (see (\ref{OmegaLSW}) and  (\ref{DiagLSW})).
Of course, this is unnecessary, since the value of $F$ can be 
deduced from the experimental data on magnon dispersion.
In terms of the unit vector $\bm U$, spontaneous magnetization
can be calculated as 
$\Lan S^z \Ran  = S \Lan U^3 \Ran \approx S - (S/2) \Lan \bm \pi^2 \Ran$,
which coincides with (\ref{SzPropagator}).
Hamiltonian (\ref{HamEffLatt0}) resembles the Hamiltonian first
obtained by 
Dyson, starting from nonorthogonal multi spin-wave states \cite{Dyson1}.
It was subsequently rederived using boson representations for the
spin operators \cite{Oguchi,Maleev}. However, 
(\ref{HamEffLatt0}) is expressed in terms of true magnon field operators,
with no direct connection to the localized spins of (\ref{HFM1}).
Also, it is seen from the derivation of (\ref{HamEffLatt0})
that $H_{\text{int}}^{(\text{II})}$ has the form of the usual gradient contribution 
in the Hamiltonian of a unit vector field, while $H_{\text{int}}^{(\text I)}$
describes magnon-magnon interactions originating  in the WZW term (A part
of interactions form WZW-term of the form $\bm \pi^2 \nabla^2 \bm \pi^2$ 
are present in $H_{\text{int}}^{(\text{II})}$ too).
athe magnon-magnon interactions collected in $H_{\text{int}}^{(\text I)}$ are
therefore essential for preserving the spin characteristics
of  bosonic field $\bm U(\bm x)$.

After the free and interaction parts of the Hamiltonian
have been constructed, the perturbation theory may be applied
to calculate the Green's function 
%-------------------------------------------------------------------------------------------------------------------
\bea
\hspace*{-0.8cm}
G(\bm x - \bm y, \tau_x - \tau_y) = 
\frac{\Lan \mbox{T} \left\{ \psi(\bm x,\tau_x) \psi^\dagger (\bm y,\tau_y) U(\beta)  \right\}  \Ran_0}
{\Lan U(\beta)\Ran_0} 
\eea  
%-------------------------------------------------------------------------------------------------------------------
with  $U(\beta) = \mbox{T} \exp \left\{  -\int_0^\beta \d \tau H_{\text{int}}(\tau) \right\}$,
and the free energy, thereby determining the influence of interaction
on magnon energies and thermodynamic properties of the system.
By expanding the exponential in definition of  $U(\beta)$, we
arrive at the Feynman rules for interacting lattice magnon fields in
O(3)  HFM.
We shall now introduce a convenient variant of Feynman diagrams.

%-------------------------------------------------------------------------------------------------------------------
\begin{figure*}
\bc 
\includegraphics[scale=0.8]{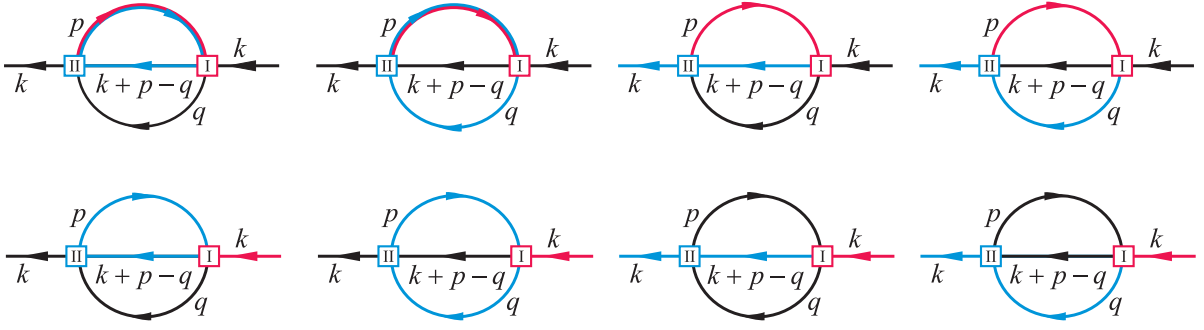}
\caption
{\label{fig4} (Color online) All possible momentum-space (colored)  contractions corresponding to the upper diagram
of FIG \ref{fig3}. Each colored line denotes the lattice Laplacian
acting on a magnon  propagator. The line carrying two colors represent propagator
affected by two Laplacians (See the text).}
\ec
\end{figure*}
%-------------------------------------------------------------------------------------------------------------------

\subsection{Diagrammar}\label{Diagrammar}

To define graphical calculations suitable for particular 
interaction in (\ref{HamEffLatt0}), i.e. in (\ref{HamEffLatt1}),
consider the two-loop diagram for self energy depicted at FIG \ref{fig3},
and a typical contraction proportional to (we abbreviate $\psi (\bm x,\tau_x)$ as  $\psi_x$, etc.)
%-------------------------------------------------------------------------------------------------------------------
\begin{widetext}
\bea
%\int_{a(\beta),b(\beta)}
v_0^2 \sum_{\bm a, \bm b} \int_{0}^\beta \d \tau_a \int_{0}^\beta \d \tau_b\;
 \Lan 0 | \text{T} \{     
\Wwick{23}{\psi_{{x}}   \;\;  \psi^\dagger_a  \;\;  \psi_a  \;\; {\nabla}^2_a ( <2{\psi^\dagger}_a  \;\; <1\psi_a   )  \;\;
\psi^\dagger_b  \;\; \psi_b \;\;  >1{\psi^\dagger}_b  \;\;  \nabla^2_b  >2\psi_b   \;\;  \psi_y^\dagger  }
{333}{<3 \psi_{{x}}   \;\;  >3\psi^\dagger_a  \;\;  <4\psi_a  \;\; {\nabla}^2_a \left( \psi^\dagger_a  \;\; \psi_a   \right)  \;\;
>4\psi^\dagger_b  \;\; <5\psi_b \;\;  \psi^\dagger_b  \;\;  \nabla^2_b  \psi_b   \;\; >5 \psi_y^\dagger }
\}  |0  \Ran . \label{Contr}
\eea 
\end{widetext}
%-------------------------------------------------------------------------------------------------------------------
Two lattice Laplacians, acting upon
magnon propagators [which are explicitly  given in (\ref{Prop})], appear in the upper integral. 
The expressions containing discrete Laplacians
could be rewritten as the difference between value of
lattice fields on a given site and on all of its nearest neighbors
(see (\ref{OpIzLapA})), which would seemingly simplify
expressions like (\ref{Contr}). 
However, the eigenvalues of $\nabla^2$ are proportional
 to the free magnon energies and
the physical interpretation favors 
 the use of lattice Laplacian.
Therefore, we will stick to  the form
explicitly containing lattice Laplacian.
To distinguish between
 two  or more Laplacians in diagrams, we introduce colored lines
and vertices. Each vertex carries  single color (red or blue
in our example), representing a single Laplacian 
contained in it. 
The lines could be single- or multi-color valued, depending
on weather one or more Laplacians acts upon them.
The rest of the lines are simply black. 
All lines are labeled by $D+1$ momentum $k = [\bm k, \omega_n]^{\text T}$
and colored ones also carry eigenvalue $-\widehat{\bm k}^2$.
The standard momentum-conservation rules at vertices
equally apply for black and colored lines.
Note that the Laplacian of $H_{\text{int}}^{(\text I)}$ always acts on a
single propagator. Thus, only single colored line, with the 
color of the propagator being the same as that of vertex,  can
end in $\begin{array}{r}
 {\includegraphics[scale=1.0]{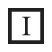}}
\end{array}$. However, it can be single or multi color-valued,
depending on if it is affected by a Laplacian of 
 another vertex. In contrast, a single
colored line of the  same color
as that of the vertex is passing  through 
$\begin{array}{r}
 {\includegraphics[scale=1.0]{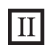}}
\end{array}$, and it carries eigenvalue $-\widehat{\bm k}^2$ of the algebraic sum
of the incoming and outgoing momenta.
The other three lines attached to $\begin{array}{r}
 {\includegraphics[scale=1.0]{Fig3_a.eps}}
\end{array}$, as well as the remaining two of $\begin{array}{r}
 {\includegraphics[scale=1.0]{Fig3_b.eps}}
\end{array}$
could be colored  differently than the vertex or  be black.
For example, the momentum-space representation of integral (\ref{Contr})
is given at the bottom of FIG \ref{fig3} and the corresponding
integrand is proportional to $\widehat{\bm k - \bm q}^{\;2} \widehat{\bm p}^{\;2}$.

The full consistency of Feynman diagrams with colored propagators
is achieved by  supplementing the rules of preceding paragraph  with additional
conventions concerning loops closed around a single
vertex.
These  appear, for
example, in the one-loop corrections to
the magnon propagator as well as in the perturbative
corrections to the free energy.
Consider first the one-loop
diagrams.
If the Laplacian of $H_{\text{int}}^{(\text{I})}$
acts on a single propagator, the loop will be 
drawn half-colored, so that only one colored 
part of the loop ends at $\begin{array}{r}
 {\includegraphics[scale=1.0]{Fig3_a.eps}}
\end{array}$. If the same situation occurs with 
$\begin{array}{r}
 {\includegraphics[scale=1.0]{Fig3_b.eps}}
\end{array}$, the line is in full color. Thus we have
%------------------------------------------------------------------------------------------------------------------------
\bea
\begin{array}{l}
\includegraphics[scale=0.7]{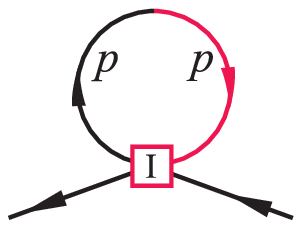}
\end{array}
\propto \frac{1}{\beta} \sum_{n = -\infty}^{\infty} \int_{\bm p} 
D(\bm p,\omega_n) \widehat{\bm p}^{\; 2}  
\eea
%------------------------------------------------------------------------------------------------------------------------
but, also
%------------------------------------------------------------------------------------------------------------------------
\bea
\hspace*{-0.8cm}\begin{array}{l}
\includegraphics[scale=0.7]{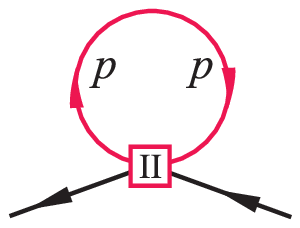}
\end{array}
\propto \frac{1}{\beta} \sum_{n = -\infty}^{\infty} \int_{\bm p} 
D(\bm p,\omega_n) \widehat{\bm p - \bm p}^{\; 2} = 0
\label{H2OneLoop} 
\eea
%------------------------------------------------------------------------------------------------------------------------
with $D(\bm p,\omega_n)$ denoting the Fourier components of 
lattice magnon propagator (\ref{Prop}).
These rules also apply to the diagrams with two loops
attached to $\begin{array}{r}
 {\includegraphics[scale=1.0]{Fig3_a.eps}}
\end{array}$. Also, they hold if the Laplacian 
of $H_{\text{int}}^{(\text{II})}$ acts on propagators
belonging to the same loop, as in (\ref{H2OneLoop}).
If, hoverer, the Laplacian of $H_{\text{int}}^{(\text{II})}$
affects propagators from different loops, they are to be 
drawn half-colored. For example
%------------------------------------------------------------------------------------------------------------------------
\bea
\hspace*{-0.2cm} \begin{array}{l}
\includegraphics[scale=0.7]{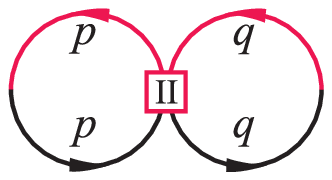}
\end{array}
\propto  \hspace{-0.1cm}  \frac{1}{\beta^2}   \hspace{-0.1cm}  \sum_{n,m = -\infty}^{\infty} \int_{\bm p,\bm q} 
\hspace{-0.1cm}  D(p) D(q) \widehat{\bm p - \bm q}^{\; 2} 
\label{H2TwoLoop} %\begin{array}{l}
\eea
%------------------------------------------------------------------------------------------------------------------------
with $p = [\bm p, \omega_n]^{\text T}, q=[\bm q, \omega_m]^{\text T}$.
Of course, it is of no importance if the upper or the lower part of
 diagram in (\ref{H2TwoLoop}) is colored. Hoverer,
both of these are not to be counted, since they represent the
same contraction.

As an example, in FIG \ref{fig4} we give the full set
of colored momentum space diagrams corresponding to the
upper diagram of FIG \ref{fig3}.
Each of these diagrams is to be multiplied by factor 2, due to
two identical sets of contractions generated by two
$\psi^\dagger_b$ operators of $H_{\text{int}}^{(\text{II})}$.
We shall refer to the diagrams with colored lines and vertices  as colored contractions.

In a final remark, we note that extension of multi-color line
formalism to higher order interactions is straightforward. A glance
on (\ref{Lint}) reveals that all vertices of effective interaction,
regardless on the number of $\pi^a$ fields, 
carry single discrete Laplacian.
Also, the method of multi-color Feynman diagrams is,
with minimal interventions, applicable to various theories 
of scalar fields with derivative couplings.
In particular, we shall find them very useful in the
Section \ref{RotSymLagr}.

%===============================================================================================================================================
\section{Magnon  self-energy at  two-loop} \label{SE2Loop}
%===============================================================================================================================================

%-------------------------------------------------------------------------------------------------------------------
\begin{figure}
\includegraphics[scale=1.0]{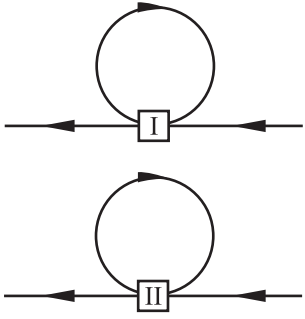}
\caption
{\label{fig1}One-loop corrections to
the lattice propagator (\ref{Prop}). The numbers associated 
with the  vertices refer to $H_{\text{int}}^{(\text I)}$ and
$H_{\text{int}}^{(\text{II})}$ of Eq. (\ref{HamEffLatt1}).}
\end{figure}
%-------------------------------------------------------------------------------------------------------------------

\subsection{One-loop correction to the magnon self-energy}

The graphs occurring at the one-loop approximation are given in FIG \ref{fig1}.
The explicit form of the correction arising from the first vertex is
easily found using Feynman rules defined above:
%------------------------------------------------------------------------------------------------------------------------
\bea
\hspace*{-0.3cm}\Sigma_{\text I}^{(1)}(\bm k,\omega_n) &=& \hspace{-0.2cm} \begin{array}{l}
\vspace{0.8cm} \includegraphics[scale=0.7]{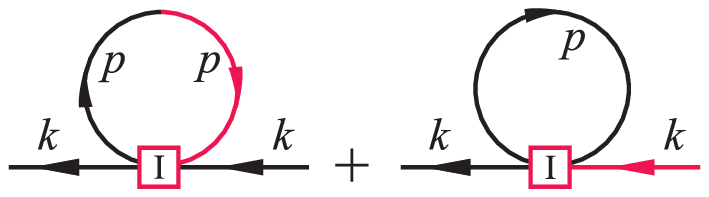}
\end{array}.  %\begin{array}{l}
\label{Diag11}
\eea
%------------------------------------------------------------------------------------------------------------------------
It is understood
that external legs, black and  colored,  are to be amputated. Further,
%------------------------------------------------------------------------------------------------------------------------
\bea
\hspace*{-0.3cm}\Sigma_{\text{II}}^{(1)}(\bm k,\omega_n) &=& \hspace{-0.2cm} 
\begin{array}{l}
\vspace{0.8cm} \includegraphics[scale=0.7]{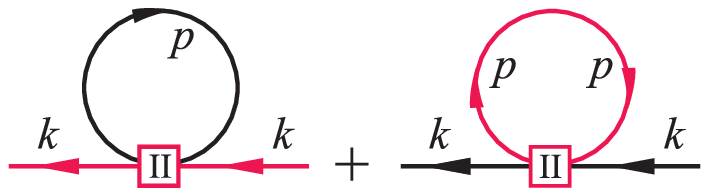} 
\end{array}\nonumber \\
& + & \hspace{-0.2cm}  \begin{array}{l}
\vspace{0.8cm} \includegraphics[scale=0.7]{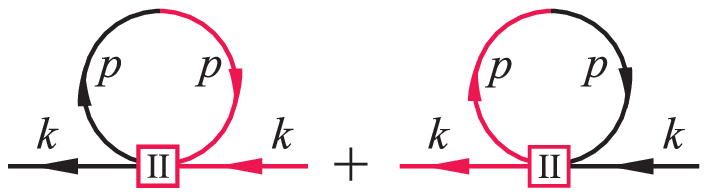} 
\end{array} %\begin{array}{l}
\label{Diag12}.  
\eea
%------------------------------------------------------------------------------------------------------------------------
\vspace*{-0.8cm}

\noindent
Since the first two diagrams of (\ref{Diag12}) vanish,
by performing summation over the Matsubara frequencies, we obtain
%-------------------------------------------------------------------------------------------------------------------
\bea
\Sigma^{(1)}(\bm k) & = & \Sigma_{\text{I}}^{(1)}(\bm k) + \Sigma_{\text{II}}^{(1)}(\bm k) \label{Sigma1} \\
& = & \frac{\widehat{\bm k}^{\;2}}{2 m_0} \; \frac{1}{S} \frac{|\bm \lambda|^2}{2 D} v_0 \int_{\bm q} \Lan n_{\bm q}
 \Ran_0\;  \widehat{\bm q}^{\;2} \nonumber
\eea
%-------------------------------------------------------------------------------------------------------------------
where  we have exploited cubic symmetry of the lattice, and 
$\Lan n_{\bm q} \Ran_0$ represents the Bose distribution
for free magnons. According to (\ref{Sigma1}), magnons acquire
mass
%-------------------------------------------------------------------------------------------------------------------
\bea
[m_{\text r}^{(1)}]^{-1} & = & m_0^{-1} 
\left[ 1-\frac 1 S \frac{|\bm \lambda|^2}{2D} v_0 \int_{\bm q} \Lan n_{\bm q} \Ran_0 \widehat{\bm q}^{\;2}
\right] \nonumber \\
& \equiv & m_0^{-1}[1-A(T)]  \label{Mr1}
\eea
%-------------------------------------------------------------------------------------------------------------------
This result can be made self-consistent bu further summation, i. e. by replacing the
propagators with full Green's functions in (\ref{Diag11}) and (\ref{Diag12})
%-------------------------------------------------------------------------------------------------------------------
\bea
[m_{\text R}^{(1)}]^{-1} = m_0^{-1} 
\left[ 1-\frac 1 S \frac{|\bm \lambda|^2}{2D} v_0 \int_{\bm q} \Lan n_{\bm q} \Ran \widehat{\bm q}^{\;2}
\right], \label{MR1}
\eea
%-------------------------------------------------------------------------------------------------------------------
with $\Lan n_{\bm q} \Ran$ denoting the Bose distribution for
magnons with energies $\widehat{\bm q}^2/(2 m_{\text{R}}^{(1)})$.
If constant $F^2$ is chosen so that $H_0$ of (\ref{HamEffLatt0})
fully coincides with (\ref{DiagLSW}), i.e. 
%-------------------------------------------------------------------------------------------------------------------
\bea
F^2 = \frac{\Sigma}{2 m_0} = \frac{J S^2 Z_1 |\bm \lambda|^2}{2 D v_0} \label{FsqDef}
\eea
%-------------------------------------------------------------------------------------------------------------------
then $\widehat{\bm k}^2/(2 m_{\text{R}}^{(1)})$ is precisely
renormalized spin-wave energy, obtained for the first time 
in \cite{MBloch} by minimization of the free energy of a ferromagnet, 
where the spin Hamiltonian (\ref{HFM1}) is written in terms
of Dyson-Maleev (DM) bosons with  only diagonal
part of the interaction being retained.
It was also obtained by the bubble diagram summation \cite{Loly}, 
again using DM representation.
However, only the derivation of 
(\ref{MR1}) using effective Lagrangian clearly shows that the effects of two distinct
types of magnon-magnon interactions  are accounted for in (\ref{MR1}).

\subsection{One-loop approximations for the spontaneous magnetization}

The one-loop corrections to the LSWT result for spontaneous
magnetization are  found by substituting magnon propagator
with Green's function calculated to the one loop in (\ref{SzPropagator}).
This is easily obtained by keeping the external legs in (\ref{Diag11})
and (\ref{Diag12}). The result is
%-------------------------------------------------------------------------------------------------------------------
\bea
\Lan S^z \Ran & = &  S - v_0 \int_{\bm p} \Lan n_{\bm p} \Ran_0  \nonumber + \delta \Lan S^z \Ran, \\
\delta \Lan S^z \Ran & = & - 
 v_0 \int_{\bm p} \frac{ \Sigma^{(1)}(\bm p)\Lan n_{\bm p} \Ran_0 [\Lan n_{\bm p} \Ran_0 +1]}{T}.
\label{OneLoopMagn}
\eea
%-------------------------------------------------------------------------------------------------------------------

There is an obvious virtue in writing the spontaneous magnetization as in Eq. (\ref{OneLoopMagn}).
The term $S-v_0\int_{\bm q} \Lan n_{\bm q} \Ran_0$ describes the reduction of 
spontaneous magnetization due to free magnons. Its low-temperature 
expansion for 3D HFM contains well known contributions proportional to $T^{3/2} $(Bloch's law), $T^{5/2}$,
$T^{7/2}$ and so on. On the other hand, the
corrections arising from  the magnon-magnon interactions are entirely
collected  in the  integral proportional to 
$1/T$. More generally, the low-temperature series for spontaneous magnetization
of a $D-$dimensional simple cubic HFM
in the one-loop approximation consists of two parts
%-------------------------------------------------------------------------------------------------------------------
\bea
\Lan S^z \Ran & = & S + \delta \Lan S^z \Ran_{\text{free}} 
+ \delta \Lan S^z \Ran_{\text{int}},
\eea
%-------------------------------------------------------------------------------------------------------------------
where
%-------------------------------------------------------------------------------------------------------------------
\bea
\delta \Lan S^z \Ran_{\text{free}} & = &  \alpha_0 T^{D/2} + \alpha_1 T^{(D+2)/2} +
\alpha_{2} T^{(D+4)/2}    \nonumber \\
& + &   \alpha_3 T^{(D+6)/2} + \O\left(T^{(D+8)/2}\right) \label{DeltaSzFreee}
\eea
%-------------------------------------------------------------------------------------------------------------------
and the free-magnon coefficients  $\alpha_i$ are given by
%-------------------------------------------------------------------------------------------------------------------
\bea
\alpha_0 & = & -\left( \frac{1}{2\sqrt{\pi}} \right)^{D} \frac{\sqrt{\pi}}{\Gamma(D/2)}
\zeta \left(\frac D2 \right)
\left[\frac{\Sigma a^2}{F^2} \right]^{D/2},
 \nonumber \\
\alpha_1 & = & -\left( \frac{1}{2\sqrt{\pi}} \right)^{D} \frac{D }{16}\;
\zeta \left(\frac{D+2}{2} \right) \left[ \frac{\Sigma a^2}{F^2}  \right]^{\frac{D+2}{2}}, \nonumber \\
\alpha_2 & = & -\left( \frac{1}{2\sqrt{\pi}} \right)^{D} \frac{D\left[ D + 8 \right]}{512}
\zeta \left(\frac{D+4}{2} \right) 
\left[ \frac{\Sigma a^2}{F^2}  \right]^{\frac{D+4}{2}},  \nonumber \\
\alpha_3 & = &- \left( \frac{1}{2\sqrt{\pi}} \right)^{D} \frac{D }{3072}\;
\zeta \left(\frac{D+6}{2} \right) \left[ \frac{\Sigma a^2}{F^2} \right]^{\frac{D+6}{2}} \nonumber \\
& \times & \left[25+3D+\frac{D^2}{8}\right]. \label{Alphai}
\eea
%-------------------------------------------------------------------------------------------------------------------
The temperature expansion of one-loop correction to LSWT results is
%-------------------------------------------------------------------------------------------------------------------
\bea
\delta \Lan S^z \Ran_{\text{int}} & = &  \beta_1 T^{D+1} + \beta_2 T^{D+2} +
 \O \left(T^{D+3}\right), \label{OneLoopCorr}
\eea
%-------------------------------------------------------------------------------------------------------------------
with
%-------------------------------------------------------------------------------------------------------------------
\bea
\beta_1 & = & -\frac{1}{S} \left( \frac{1}{2 \pi} \right)^{2D} \frac{D \pi^D}{8}
\left[\frac{\Sigma a^2}{F^2} \right]^{D+1} \nonumber \\
&\times& \zeta \left(\frac D2 \right) \zeta \left(\frac{D+2}{2} \right),
 \nonumber \\
\beta_2 & = & - \frac 1S \left( \frac{1}{2 \pi} \right)^{2D} \frac{D [D+2] \pi^D}{128}
 \left[\frac{\Sigma a^2}{F^2} \right]^{D+2}
\nonumber \\
& \times & \left\{  \left[  \zeta \left(\frac{D+2}{2} \right) \right]^2 + 
\zeta \left(\frac{D}{2} \right) \zeta \left(\frac{D+4}{2} \right)   \right\}. \label{Betai}
\eea
%-------------------------------------------------------------------------------------------------------------------
In the formulae above, $\zeta(x)$ denotes the Riemann zeta function and 
$D \geq 3$ is understood. 
%-------------------------------------------------------------------------------------------------------------------
\begin{figure}
\bc 
\includegraphics[scale=0.8]{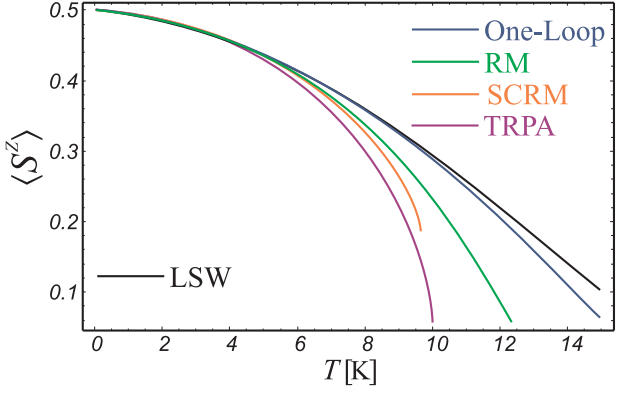}
\caption
{\label{fig5}(Color online) Spontaneous
magnetization of $S = 1/2$ and $J = 10$K 3D HFM calculated
using LSWT (Eq. (\ref{SzPropagator})), Tyablikov RPA,
one-loop approximation (\ref{OneLoopMagn}),
renormalized magnons (RM) of (\ref{Mr1}) and self-consistent renormalized 
magnons (SCRM) given in (\ref{MR1}).}
\ec
\end{figure}
%-------------------------------------------------------------------------------------------------------------------
For $D=3$, the lowest
order correction from magnon-magnon interaction comes to be $\propto T^4$, in 
agreement with Dyson \cite{Dyson2}. 
We note that the correct form of leading order contribution is found  easily, 
evaluating only a single type of diagram indicated at FIG \ref{fig1}. 
This should be compared with  continuum field-theoretical calculations
\cite{Hofmann2,Hofmann3}, where number of diagrams to be 
evaluated becomes greater with increasing
dimensionality of the lattice (see also the Section \ref{RotSymLagr}).
Also, the lattice regularized theory allows for  a comparison   with LSWT and
other methods, such as Tyablikov RPA,
even at not too low temperatures.

The plot of spontaneous magnetization of spin $S=1/2$ 
and exchange integral $J = 10$K calculated by LSWT (Eq. (\ref{SzPropagator})),
Tyablikov RPA, the one-loop approximation of Eq. (\ref{OneLoopMagn})
and dressed magnons of (\ref{Mr1}) and (\ref{MR1})
is presented at FIG. \ref{fig5}. We have set $F^2 = J S \Sigma a^2$ in (\ref{OneLoopMagn}) to work with
a common energy scale.
The TRPA result for $T_{\text C}$ is $\approx 10.065$ K
(For precise calculation of the critical temperature in TRPA, see
\cite{SSCTNRPA,JPA} and references therein).

\subsection{Two-loop corrections to the self-energy}

\begin{figure}
\bc 
\includegraphics[scale=0.8]{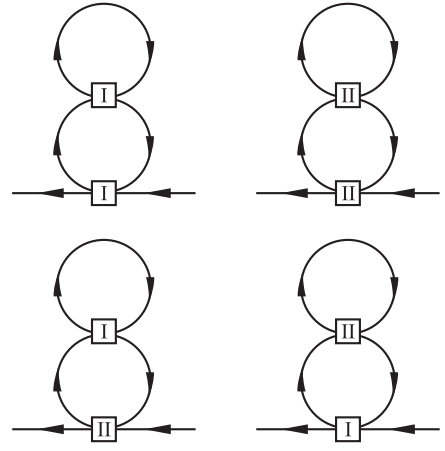}
\caption
{\label{fig8}Two-loop graphs for magnon self-energy that contribute solely to the
magnon mass.}
\ec
\end{figure}

The self-energy graphs with two loops, involving vertices 
$\begin{array}{r}
 {\includegraphics[scale=1.0]{Fig3_a.eps}}
\end{array}$ and $\begin{array}{r}
 {\includegraphics[scale=1.0]{Fig3_b.eps}}
\end{array}$ can be classified in two groups.
Graphs presented at FIG \ref{fig8} contribute 
purely to the magnon mass, i.e. they have no imaginary parts.
%-------------------------------------------------------------------------------------------------------------------
%-------------------------------------------------------------------------------------------------------------------
%-------------------------------------------------------------------------------------------------------------------
\begin{figure}
\bc 
\includegraphics[scale=0.8]{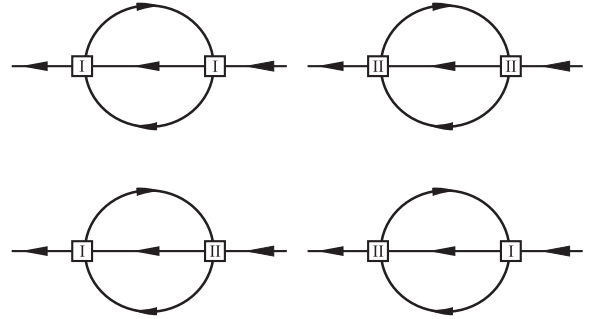}
\caption
{\label{fig9}Two-loop diagrams for magnon self-energy that
cause finite magnon lifetime.}
\ec
\end{figure}
%-------------------------------------------------------------------------------------------------------------------
The diagrams from FIG \ref{fig9}, however, produce the finite magnon
lifetime. (The influence of 
finite magnon lifetime shall not be discussed further in the
present paper. For details and references, 
see \cite{KaganovChubukov,Harris,Zitomirski}.) All these diagrams can be  evaluated 
using formalism of colored propagators, as explained in
\ref{Diagrammar}. An example for decomposition of compact
two-loop diagram of FIG \ref{fig9} into its momentum-space contractions is
given at FIG \ref{fig4}. 
Note that there are only two distinct 
Matsubara summations at the two-loop. The first one 
is common to all graphs from FIG \ref{fig8} and the 
second one appears in all graphs form FIG \ref{fig9}, 
since the lattice Laplacian leaves $n$-index
untouched.

\begin{widetext}
The diagrams from FIG \ref{fig8} then evaluate to 
%-------------------------------------------------------------------------------------------------------------------
\bea
\Sigma_{\text{FIG 5}}(\bm k) & = &  \frac{1}{S} \frac{\widehat{\bm k}^2}{2 m_0 }
\frac{2D}{|\bm \lambda|^2} \; \frac{\alpha(T)}{2 m_0 T} 
\left[\frac{|\bm \lambda|^2}{2 D}\right]^2  v_0 \int_{\bm p} \frac{ \Lan n_{\bm p} \Ran_0 [\Lan n_{\bm p} \Ran_0 +1]}{T} 
\left[ \widehat{\bm p}^{\;2}  \right]^2 \equiv \frac{\widehat{\bm k}^2}{2 m_0 } A(T) B(T)
\label{TwoLoopSE1}
\eea
%-------------------------------------------------------------------------------------------------------------------
The contribution from diagrams presented at FIG \ref{fig9}
consist of two parts, both of which change the
geometry of magnon dispersion: The first part is proportional to 
$\widehat{\bm k}^2$
%-------------------------------------------------------------------------------------------------------------------
\bea
\Sigma_{\text{FIG 6}}^{(a)}(k) & = &  \frac{1}{2 S^2} \frac{\widehat{\bm k}^2}{[2 m_0 ]^2}
\left[  \frac{|\bm \lambda|^2}{2 D}  \right]^2 v_0^2 \int_{\bm p,\bm q}
F^{\bm k}_{\bm p, \bm q}(\i \omega_n) 
 \widehat{\bm q}^{\;2} \widehat{\bm p}^{\;2} 
\left[ \widehat{\bm q}^{\;2}   - \widehat{\bm p - \bm q}^{\;2} \right]
\label{TwoLoopSE2}
\eea
%-------------------------------------------------------------------------------------------------------------------
and the other one  to $\widehat{\bm k}^4$
%-------------------------------------------------------------------------------------------------------------------
\bea
\Sigma_{\text{FIG 6}}^{(b)}(k) & = &  \frac{1}{2 S^2} \left[ \frac{\widehat{\bm k}^2}{2 m_0 }\right]^2
\left[  \frac{|\bm \lambda|^2}{2 D}  \right]^2 v_0^2 \int_{\bm p,\bm q}
F^{\bm k}_{\bm p, \bm q}(\i \omega_n) 
\widehat{\bm q}^{\;2} 
\left[ \widehat{\bm q}^{\;2}   - \widehat{\bm p - \bm q}^{\;2} \right].
\label{TwoLoopSE3}
\eea
%-------------------------------------------------------------------------------------------------------------------
 \noindent
We have introduced here a shorthand notation for the vertex function,
obtained by double Matsubara-index summation
%-------------------------------------------------------------------------------------------------------------------
\bea
F^{\bm k}_{\bm p, \bm q}(\i \omega_n)  \label{Fkpg} 
 = \frac{
\Lan n_{\bm p} \Ran_0 [1+\Lan n_{\bm q} \Ran_0 + \Lan n_{\bm k + \bm p - \bm q} \Ran_0]
-\Lan n_{\bm q} \Ran_0 \Lan n_{\bm k + \bm p - \bm q} \Ran_0}
{\omega_{0}(\bm k + \bm p - \bm q)- \omega_{0}(\bm p) + \omega_{0}(\bm q) - \i \omega_n}.
\nonumber
\eea
%-------------------------------------------------------------------------------------------------------------------
From (\ref{Sigma1}), (\ref{TwoLoopSE1}), (\ref{TwoLoopSE2}) and (\ref{TwoLoopSE3}),
we find the magnon energies at two loop
%-------------------------------------------------------------------------------------------------------------------
\bea
\omega_{\text{2loop}}(\bm k) & = & \omega_0(\bm k) - \delta\omega_{\text{2loop}}(\bm k), 
\hspace{1cm}  \delta\omega_{\text{2loop}}(\bm k) =   \lim_{\delta \rightarrow 0}      
\text{Re} \Sigma(\bm k, \omega_{0}(\bm k)+ \i \delta),   \label{deltaOmega2Loop} \\
\Sigma(\bm k, \omega_{0}(\bm k)+ \i \delta) & = & 
 \Sigma^{(1)}(\bm k) + \Sigma_{\text{FIG 5}}(\bm k) + \Sigma_{\text{FIG 6}}^{(a)}\left(\bm k,  \omega_0(\bm k) + \i \delta \right)
+\Sigma_{\text{FIG 6}}^{(b)}\left(\bm k,  \omega_0(\bm k) + \i \delta \right) \nonumber \\
& = & \frac{\widehat{\bm k}^2}{2 m_0} \left[
A(T) + A(T) B(T)\right] \nonumber \\
&+& \frac{\widehat{\bm k}^2}{2 m_0}  \frac{1}{2 S^2} 
\left[  \frac{|\bm \lambda|^2}{2 D}  \right]^2 \frac{v_0^2}{2 m_0 } \int_{\bm p,\bm q}
F^{\bm k}_{\bm p, \bm q}\left(\omega_0(\bm k)+\i \delta \right) 
 \widehat{\bm q}^{\;2}\left( \widehat{\bm p}^{\;2} + \widehat{\bm k}^{\;2} \right)
\left( \widehat{\bm q}^{\;2}   - \widehat{\bm p - \bm q}^{\;2} \right).
  \label{2LoopOmega}
\eea
%-------------------------------------------------------------------------------------------------------------------
\end{widetext}
It is seen from (\ref{2LoopOmega}) that magnons remain gapless at two loop 
($\omega_{\text{2loop}}(\bm k) \rightarrow 0$ as $|\bm k| \rightarrow 0$)
just as do pions in Lorentz-invariant models \cite{Smilga,Lewis,StrangeMass}.

\begin{figure}
\bc 
\includegraphics[scale=0.7]{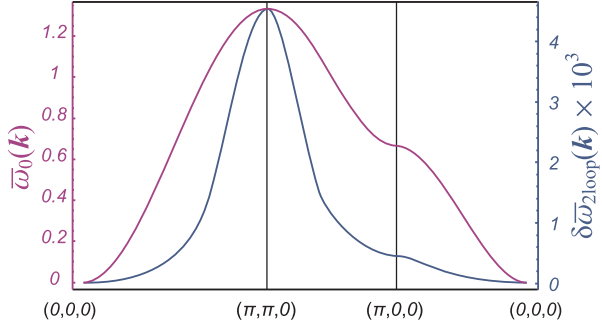}
\caption
{\label{figMD}(Color online) Reduced free magnon energies 
$\bar{\omega}_0(\bm k) = \omega_0(\bm k) [m_0 |\bm \lambda|^2/D]$ 
(left axis, purple curve)
and  2-loop correction $\delta \bar{\omega}_{\text{2loop}}(\bm k) =
 \delta \omega(\bm k)_{\text{2loop}}  [m_0 |\bm \lambda|^2/D]$ (right axis, blue curve).
$\omega_0(\bm k)$ is defined in (\ref{OmegaLSW}),  and
$ \delta \omega_{\text{2loop}}(\bm k)$ is given
in  (\ref{deltaOmega2Loop}) and (\ref{2LoopOmega}). $S=1/2, D=3 J=10 \text{K}$ and $T=1$K for both curves}
\ec
\end{figure}

On FIG \ref{figMD} we plot the free-magnon dispersion $\omega_0(k_x,k_y,0)$ [Equation (\ref{OmegaLSW})]
for $J=10$K, $S=1/2$ and $T=1$K,
along with $\delta \omega_{\text{2loop}}(k_x,k_y,0)$. The 
numerical values of magnon mass renormalizating factors 
are $A = 2.849 \times 10^{-4}$ and $B=7.298 \times 10^{-4}$.

\subsection{TRPA as an effective field theory} \label{EFTRPA}

Now that the picture of HFM as a interacting magnon field
 is complete, we 
can make some observation on TRPA result for spontaneous magnetization
and free energy.
They may not be apparent, or even accessible  within conventional TGF methodology
or any other approach that relies on boson/fermion representation
of spin operators.
Present derivation of TRPA dispersion relation for magnons, and latter discussion
on spurious $T^3$ term,  clearly isolates the influence of retained magnon-magnon 
interactions from the neglection of short-ranger fluctuations
in mean number of magnons per lattice site.

Consider a system of  magnons for which the free Lagrangian is (\ref{Lfree})
and interaction is described  by  $\widetilde{\L}_{\text{int}}$ obtained from (\ref{Lint})
by neglecting the second term proportional to $\bm \pi \cdot \Delta \bm \pi$ and
keeping only $\sqrt{1- \bm \pi^2}$ in the square bracket. Corresponding
Hamiltonian that includes up to four magnon operators,
 $\widetilde{H}  =  H_0 + \widetilde{H}_{\text{int}}$,
is easily constructed. Since $\widetilde{H}_{\text{int}} = -H_{\text{int}}^{(\text{II})}$,
the one-loop self energy is found from (\ref{Diag12})
%-------------------------------------------------------------------------------------------------------------------
\bea
\widetilde{\Sigma}(\bm k) = \frac{1}{S} \frac{v_0}{2 m_0} \int_{\bm p}
\Lan n_{\bm p} \Ran_0 \; \widehat{\bm p - \bm k}^2 \label{SigmaRPA}.
\eea
%-------------------------------------------------------------------------------------------------------------------
We shall now assume that the mean number of excited magnons is the same
at each lattice site. This simplification mimics  TRPA 
replacement of the operator $S_{\bm n}^z$ with the site-independent average
$\Lan S^z \Ran$.
Then 
%-------------------------------------------------------------------------------------------------------------------
\bea
  v_0   \int_{\bm p} \Lan n_{\bm p} \Ran_0
\gamma( \bm p)  =   v_0  \int_{\bm p} \Lan n_{\bm p} \Ran_0
\eea
%-------------------------------------------------------------------------------------------------------------------
equals the mean number of magnons on each lattice site, $\Lan n_{\bm x} \Ran$.
The magnon energies may now be written as
%-------------------------------------------------------------------------------------------------------------------
\bea
\widetilde{\omega}(\bm k) = \omega_0(\bm k) - \widetilde{\Sigma}(\bm k) =  \frac{\widehat{\bm k}^2}{2 m_0} 
\frac{S-\Lan n_{\bm x} \Ran }{S},
\eea
%-------------------------------------------------------------------------------------------------------------------
which we may, for low temperatures,  identify with TRPA energies (\ref{OmegaTRPA}).
In other words, the effective Hamiltonian of Tyablikov RPA, written in terms 
of lattice magnon fields is
%-------------------------------------------------------------------------------------------------------------------
\bea
\widetilde{H} \equiv H_{\text{eff}}^{\text{RPA}} & = & H_0 +  
\frac{F^2}{8}v_0
\sum_{\bm x} 
\bm \pi^2(\bm x)  \nabla^2 \bm \pi^2(\bm x)
\label{HamEffLattRPA}
\eea
%-------------------------------------------------------------------------------------------------------------------
with $H_0$ defined in (\ref{HamEffLatt0}).
Reversing the arguments that lead to the $H_{\text{eff}}^{\text{RPA}}$,
and also to the correct interacting Hamiltonian of lattice magnons (\ref{HamEffLatt1}), we see
that TRPA results are generated starting from the % manifestly $U(1)$
leading order effective Lagrangian
%-------------------------------------------------------------------------------------------------------------------
\bea
\L_{\text{eff}}^{\text{RPA}} & = & \frac \Sigma 2\left(\partial_t U^1 U^2 - \partial_t U^2 U^1\right)
-\frac{F^2}{2} \partial_{\alpha} U^i \partial_{\alpha} U^i \nonumber \\
& - & \frac{F^2}{4} \bm \pi^2(\bm x)  \Delta \bm \pi^2(\bm x). \label{LagrRPA}
\eea
%-------------------------------------------------------------------------------------------------------------------
This Lagrangian manifestly   violates  spin-rotational invariance
of the original Heisenberg Hamiltonian (\ref{HFM1}).

%-------------------------------------------------------------------------------------------------------------------
%-------------------------------------------------------------------------------------------------------------------

Various explanations for the spurious $T^3$ term   in TRPA expansion of
the ferromagnetic order parameter
and 
the error caused by the Tyablikov decoupling at low temperatures
 have been offered by many authors.
For example, in the Tyablikov's monograph, it is attributed  to the
''approximate character'' (of the decoupling approximation)
and the ''neglection of the fluctuation of order parameter'' \cite{Tjablikov}.
In the context of the spin-diagram technique, authors of \cite{VLP1,VLP2}
state that $T^3$ arises since  ''in the decoupling methods
terms after $r_0^{-3}$ are taken into the account incorrectly''.
(Here $r_0^{-3}$ represents the formal expansion parameter in
the spin-diagram technique, namely the reciprocal interaction volume.)
In  \cite{Fishman}, the main feature of TRPA  is recognized as being 
"uncontrolled expansion to all orders in $1/Z_1$".
On the other hand, the authors of \cite{Stinchcombe} conclude that 
erroneous $T^3$ term in TRPA 
''comes from taking expectation values in the equation of motion too soon''.
Finally, in \cite{Plakida}, Tyablikov RPA is described as an approximation
''in which contributions of static fluctuations
of spins are neglected''. However, it is also noted in this reference that
$T^3$ term will appear in any approximation that incorrectly
treats spectral density entering the correlation function $\Lan S^- S^+ \Ran_{\bm k}$.
All arguments quoted above rely directly  on the localized
spin operators \cite{Tjablikov,Plakida,Fishman} that define Heisenberg Hamiltonian 
(\ref{HFM1})  or on their boson/fermion
representations \cite{VLP1,VLP2,Stinchcombe}.
The derivation of Tyablikov RPA in terms of lattice magnon fields,
as given in the present paper,
provides a simple and straightforward answer based
on the internal symmetries of the Heisenberg model. 
It is seen from the equations (\ref{HamEffLattRPA})-(\ref{LagrRPA})
that Tyablikov RPA incorrectly describes O(3) HFM at low temperatures  since
it eventually results from the effective Lagrangian (\ref{LagrRPA})
that does not preserve spin-rotational symmetry of the Heisenberg
ferromagnet (\ref{HFM1}). 
Explicitly, interactions of the form $\bm \pi^2 \bm \pi  \cdot \nabla^2 \bm \pi$ arising from the WZW term
$H_{\text{int}}^{(1)}$ are omitted in TRPA.
It may also be said that due this reduction of magnon-magnon interactions in the
effective Lagrangian, localized spins
of HFM are inadequately  described by
the  unit vector $\bm U(x)$.

We note that essential error in TRPA is made
 when
magnon-magnon interactions arising from the WZW term
are omitted.  Neglection of the SRF of the 
order parameter, i.e. neglection of the fluctuations
in the mean number of magnons at adjacent sites
 (the replacement of $v_0\int_{\bm q} \Lan n_{\bm q} \Ran_0 \gamma(\bm q)$
with $v_0\int_{\bm q} \Lan n_{\bm q} \Ran_0 = \Lan n_{\bm x} \Ran$) merely 
modifies coefficient  of the $T^3$ term.
To show this, we find the first order correction to the 
spontaneous magnetization based  on Eq. (\ref{Diag12})
%-------------------------------------------------------------------------------------------------------------------
\bea
\delta \Lan S^z \Ran & = & -  \frac{1}{S} \frac{F^2}{T \Sigma} v_0 \int_{\bm p}
 J(\bm p) \Lan n_{\bm p} \Ran_0 [\Lan n_{\bm p} \Ran_0 +1], \nonumber \\
J(\bm p) & = &  v_0 \int_{\bm q} \Lan n_{\bm q} \Ran_0 (\widehat{\bm p- \bm q})^2. \label{DeltaSzRPA}
\eea
%-------------------------------------------------------------------------------------------------------------------
The leading order term in temperature expansion of   Eq. (\ref{DeltaSzRPA})
for $D$ dimensional simple cubic lattice is
%-------------------------------------------------------------------------------------------------------------------
\bea
& - & \frac{1}{S} \frac{F^2 D \pi^D}{2 \Sigma} \left( \frac{a}{2 \pi} \right)^{2D}
 \left[  \frac{\Sigma}{F^2} \right]^{D+1} \label{DeltaSzRPA2}  \\
& \times & \left\{  \left[  \zeta \left(\frac{D}{2} \right) \right]^2 + 
\zeta \left(\frac{D}{2}+1 \right) \zeta \left(\frac{D}{2} -1 \right)
   \right\} T^D \nonumber.
\eea
%-------------------------------------------------------------------------------------------------------------------
and for $D=3$, it corresponds to spurious $T^3$ term. If an additional assumption 
on the absence of SRF is included, the term 
$\propto \zeta \left(D/2+1 \right) \zeta \left(D/2 -1 \right)$  is missing from
(\ref{DeltaSzRPA2}). The Tyablikov's $T^3$ term \cite{Tjablikov, TerHaar},
%-------------------------------------------------------------------------------------------------------------------
\bea
& - & \frac{3}{2 S} \left( \frac{1}{4 \pi J S} \right)^3 
\left[  \zeta \left(\frac{3}{2} \right) \right]^2 T^3 \label{DeltaSzRPA3} 
\eea
%-------------------------------------------------------------------------------------------------------------------
is found by setting  $F^2=J S \Sigma a^2$ (See (\ref{FsqDef})).
An alternative formulation of O(2) model (\ref{HamEffLattRPA})
is given in the Appendix \ref{AppRPA}.

%===============================================================================================================================================
\section{Free energy at three-loop}  \label{3LFE}
%===============================================================================================================================================

For the subsequent analysis of the low temperature thermodynamics, 
we shall include weak external magnetic field directed along the 3-axis.
It opens the gap in magnon spectrum
%-------------------------------------------------------------------------------------------------------------------
\bea
\omega(\bm p) \longrightarrow \omega(\bm p,H) =  \omega(\bm p) + \mu H, \label{MagFieldGap}
\eea
%-------------------------------------------------------------------------------------------------------------------
and it is included in the effective Lagangian by standard Zeeman term \cite{PRD}.
%-------------------------------------------------------------------------------------------------------------------
\be
\L_{H} = \Sigma \mu H U^3.
\ee
%-------------------------------------------------------------------------------------------------------------------
Note that the interaction Hamiltonian does not contain terms
proportional to the external field $H$. This is an exact result
to all orders in $\bm \pi ^2$, and it follows from the 
equation of motion.

\vspace*{0.5cm}

\subsection{Two-loop correction to the  free energy}

The first-order correction to the free energy (see e.g. \cite{Kapusta}) of lattice magnons,
involving the two-loop graphs, 
is given by 
%-------------------------------------------------------------------------------------------------------------------
\bea
\delta F_{\text{2loop}} = -T \begin{array}{l}
%\vspace{1.1cm} 
\includegraphics[scale=0.8]{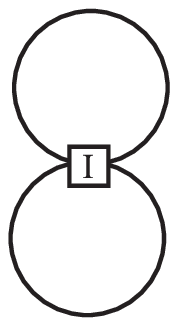}
\end{array} - T
\begin{array}{l}
%\vspace{1.1cm} 
\includegraphics[scale=0.8]{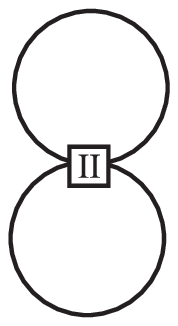}
\end{array},
\eea
%-------------------------------------------------------------------------------------------------------------------
%\vspace*{-0.5cm}
\noindent with the notation introduced in previous sections.
%-------------------------------------------------------------------------------------------------------------------
\begin{figure*}
\bc 
\includegraphics[scale=0.9]{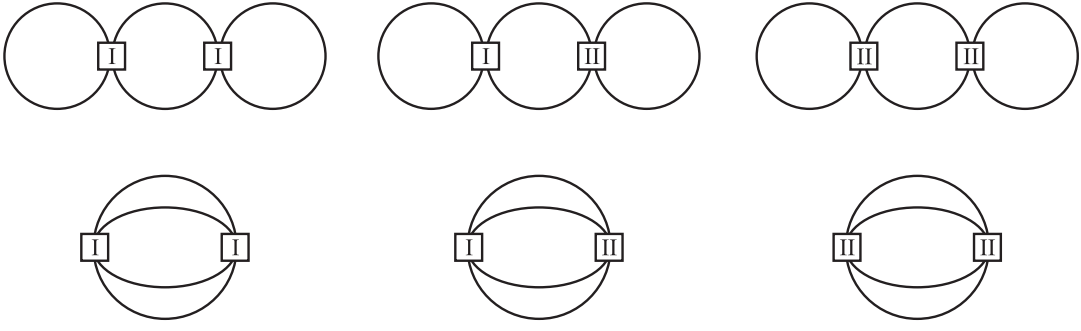}
\caption
{\label{fig7} Six distinct three-loop  diagrams contributing to the free energy.}
\ec
\end{figure*}
%-------------------------------------------------------------------------------------------------------------------
According to the Feynman rules defined in \ref{Diagrammar}, these
diagrams decompose to 
%-------------------------------------------------------------------------------------------------------------------
\bea
 \begin{array}{l}
%\vspace{1.1cm} 
\includegraphics[scale=0.8]{Fig14_a.eps}
\end{array} = 
\begin{array}{l}
%\vspace{1.1cm} 
\includegraphics[scale=0.8]{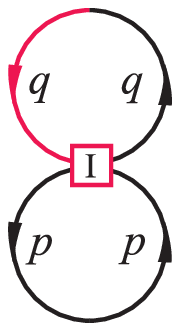}
\end{array}
+
\begin{array}{l}
%\vspace{1.1cm} 
\includegraphics[scale=0.8]{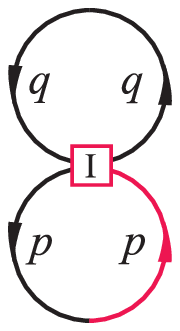}
\end{array},
\eea
%-------------------------------------------------------------------------------------------------------------------
and
%-------------------------------------------------------------------------------------------------------------------
\bea
 \begin{array}{l}
%\vspace{1.1cm} 
\includegraphics[scale=0.8]{Fig15_a.eps}
\end{array} = 
\begin{array}{l}
%\vspace{1.1cm} 
\includegraphics[scale=0.8]{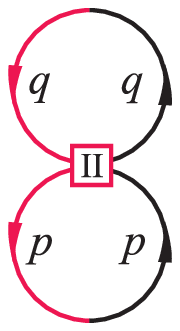}
\end{array}
+
\begin{array}{l}
%\vspace{1.1cm} 
\includegraphics[scale=0.8]{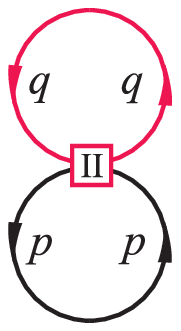}
\end{array},
\eea
%-------------------------------------------------------------------------------------------------------------------
so that the first correction to the free energy per lattice site
is given by 
%-------------------------------------------------------------------------------------------------------------------
\bea
\delta f_{\text{2loop}} = \frac{\delta F_{\text{2loop}}}{N} = - \frac{1}{S} \frac{F^2}{\Sigma}
\frac{|\bm{\lambda}|^2}{4 D} \left[ v_0 \int_{\bm p}
\Lan n_{\bm p}
 \Ran_0\; \widehat{\bm p}^{\;2} \right]^2, \label{FreeEnLatt}
\eea
%-------------------------------------------------------------------------------------------------------------------
The temperature expansion of (\ref{FreeEnLatt}) starts with
$T^{D+2}$ term. Specifically, for $D-$dimensional cubic lattices,
it is
%-------------------------------------------------------------------------------------------------------------------
\bea
\delta f_{\text{2loop}} & = & - \frac{1}{S}\frac{D \pi^D}{16 a^D} \left(\frac{1}{2 \pi} \right)^{2D}
 \left( \frac{\Sigma a^2}{F^2} \right)^{D+1}  T^{D+2} \nonumber \\
 & \times & \left[ \sum_{n = 1}^{\infty} \frac{\e^{-\mu H n /T}}{n^{(D+2)/2}} \right]^2
 +\O(T^{D+3}).
 \label{DysonFreeEnT5}
\eea
%-------------------------------------------------------------------------------------------------------------------
If $D=3$, this gives leading-order part of Dyson's $T^5$ term \cite{Dyson2}.
It receives contribution from higher-loop diagrams in the lattice
regularized theory (see the subsection \ref{3LAnalysis}).

At this point we also justify one-loop calculation of  
spontaneous magnetization from previous section
  by showing that corrections to
LSWT from (\ref{OneLoopMagn})
 can be obtained from the
first-order correction to the free energy of lattice magnons.
The first correction to spontaneous magnetization (per lattice site) is
$\delta \Lan S^z \Ran = - \partial (\delta f_{\text{2loop}})/\partial (\mu H)|_{H = 0}$.
Differentiating (\ref{FreeEnLatt}) with respect to $\mu H$ and setting $H = 0$ 
 we readily recover  equation 
 (\ref{OneLoopMagn}).

\vspace*{0.5cm}

\subsection{Three-loop corrections to the free energy}

Three-loop contribution to the magnon free energy is represented
by diagrams from FIG \ref{fig7}. 
They can be classified into two  categories, distinguished by 
$a$ and $b$ superscripts in the following equations.

Each of the three upper
diagrams  from FIG \ref{fig7}, to be classified as $a$-type, 
consists of  a number of different colored contractions. 
For the graph containing vertices solely from  $H_{\text{int}}^{(\text I)}$,
each of distinct colored contractions repeats  four times, so that
%-------------------------------------------------------------------------------------------------------------------
\bea
\hspace*{-0.5cm}\delta f_{\text{I,I}}^{(a)}& =& - 
\frac{1}{2S^2} \left[ \frac{1}{2 m_0} \right]^2 v_0^3 \int_{\bm k, \bm p, \bm q} 
\frac{  \Lan n_{\bm q}\Ran_0   [\Lan n_{\bm q}\Ran_0 +1]}{T} \nonumber \\
& \times&  \Lan n_{\bm p}\Ran_0    \Lan n_{\bm k}\Ran_0  
\left[ \widehat{\bm p}^2\widehat{ \bm q}^2  + \widehat{\bm k}^2\widehat{ \bm q}^2 
+ \widehat{\bm p}^2\widehat{ \bm k}^2 + \widehat{\bm q}^2\widehat{ \bm q}^2 \right]. \label{3loopFEI-Ia}
\eea
%-------------------------------------------------------------------------------------------------------------------
Further, for the graph with two vertices from $H_{\text{int}}^{(\text{II})}$,
we find
%-------------------------------------------------------------------------------------------------------------------
\bea
\hspace*{-0.5cm}\delta f_{\text{II,II}}^{(a)}& = & - \frac{1}{2S^2} 
\left[ \frac{1}{2 m_0} \right]^2 v_0^3 \int_{\bm k, \bm p, \bm q} 
\frac{  \Lan n_{\bm q}\Ran_0   [\Lan n_{\bm q}\Ran_0 +1]}{T} \nonumber \\
& \times&  \Lan n_{\bm p}\Ran_0    \Lan n_{\bm k}\Ran_0  
\left[ \widehat{\bm p - \bm q}^2  \widehat{\bm k - \bm q}^2 \right].\label{3loopFEII-IIa}
\eea
%-------------------------------------------------------------------------------------------------------------------
This term consists of two different colored contractions, each
appearing twice.
The total contribution of the graphs that
involve vertices of both $H_{\text{int}}^{(\text{I})}$ and $H_{\text{int}}^{(\text{II})}$
is
%-------------------------------------------------------------------------------------------------------------------
\bea
\hspace*{-0.5cm}\delta f_{\text{I,II}}^{(a)}& = &  \frac{1}{S^2}
\left[ \frac{1}{2 m_0} \right]^2 v_0^3 \int_{\bm k, \bm p, \bm q} 
\frac{  \Lan n_{\bm q}\Ran_0   [\Lan n_{\bm q}\Ran_0 +1]}{T} \nonumber \\
& \times&  \Lan n_{\bm p}\Ran_0    \Lan n_{\bm k}\Ran_0  
\left[ \widehat {\bm p}^2 \; 
\widehat{\bm k - \bm q}^2 + \widehat {\bm q}^2 \; \widehat{\bm k - \bm q}^2 \right].
\label{3loopFEI-IIa}
\eea
%-------------------------------------------------------------------------------------------------------------------
In this case, one of the colored contractions appears eight times
and the other two four times each.

\begin{widetext}
Finally, by putting contributions (\ref{3loopFEI-Ia})-(\ref{3loopFEI-IIa}) together, we find the
first part of three-loop contribution to the magnon free-energy per lattice site 

%-------------------------------------------------------------------------------------------------------------------
\bea
\delta f_{\text{3loop}}^{(a)}& =& - \frac{1}{2S^2} \left[ \frac{1}{2 m_0} \right]^2 v_0 \int_{\bm q} 
\frac{  \Lan n_{\bm q}\Ran_0   [\Lan n_{\bm q}\Ran_0 +1]}{T} 
 \left[v_0 \int_{\bm p} \Lan n_{\bm p}\Ran_0   
\left( \widehat{\bm p - \bm q}^2 - \widehat{\bm p}^2 - \widehat{\bm q}^2  \right)\right]^2.
\label{3LoopFEa}
\eea
%-------------------------------------------------------------------------------------------------------------------

The calculation resumes in a similar manner for remaining diagrams of $b$-type, 
i.e. for lower three diagrams of FIG \ref{fig7}. They add up to 

%-------------------------------------------------------------------------------------------------------------------
\bea
\delta f_{\text{3loop}}^{(b)}& =&  -\frac{1}{8 S^2}\left[ \frac{1}{2 m_0} \right]^2
v_0^3 \int_{\bm k, \bm p, \bm q} \hspace{-0.2cm}  G^{\bm q}_{\bm k. \bm p}
\left[ (\widehat{\bm k + \bm p - \bm q})^2  + \widehat{\bm q}^2 - \widehat{\bm p - \bm q}^2
- \widehat{\bm k - \bm q}^2 \right] 
 \left[  \widehat{\bm p}^2   + \widehat{\bm k}^2  - \widehat{\bm p - \bm q}^2
  - \widehat{\bm k - \bm q}^2  \right]
\label{3loopFEb}
\eea
%-------------------------------------------------------------------------------------------------------------------
with three-index Matsubara sum 
%-------------------------------------------------------------------------------------------------------------------
\bea
G^{\bm q}_{\bm k, \bm p}   \label{Gqkp} 
 = \frac{
\Lan n_{\bm k} \Ran_0 \Lan n_{\bm p} \Ran_0 [1+\Lan n_{\bm q} \Ran_0 + \Lan n_{\bm k + \bm p - \bm q} \Ran_0]
-\Lan n_{\bm q} \Ran_0 \Lan n_{\bm k + \bm p - \bm q} \Ran_0
[\Lan n_{\bm k} \Ran_0 + \Lan n_{\bm p} \Ran_0 + 1]}
{\omega_{0}(\bm k + \bm p - \bm q) + \omega_{0}(\bm q) - \omega_{0}(\bm p)  - \omega_{0}(\bm k)}.
\nonumber
\eea
%-------------------------------------------------------------------------------------------------------------------
\end{widetext}

Now we  consider limit $a\rightarrow 0$ and examine three-loop 
integrals (\ref{3LoopFEa}) and (\ref{3loopFEb}) in detail.

\subsection{Analysis of the three-loop integrals} \label{3LAnalysis}

First, it is easily seen that $\delta f_{\text{3loop}}^{(a)}$ vanishes in continuum limit,
thus contributing nothing when the lattice anisotropies are neglected. 
This was 
pointed out in \cite{Hofmann3} for $D = 3$.
Its lowest order
contribution is proportional to $T^{[3 D+6]/2}$. 
For $D$ dimensional simple cubic lattice, this  term is
%-------------------------------------------------------------------------------------------------------------------
\bea
%\delta f_{\text{3loop}}^{(a)} 
& - & \frac{v_0}{\Sigma^2} \left( \frac{\Sigma}{F^2}  \right)^{(3 D+4)/2} 
\left( \frac{a^2}{2 D}  \right)^2 \frac{D^3 [D+2]}{32} \nonumber \\
& \times & \left( \frac{1}{2 \sqrt{\pi}} \right)^{3D} \left[ \zeta\left( \frac{D+2}{2}  \right) \right]^3
T^{[3 D+6]/2}.   \label{3LoopZero}
\eea
%-------------------------------------------------------------------------------------------------------------------
It turns out, however, that this 
is not the leading order correction to the Dyson's $T^{D+2}$ term.

The leading order correction to $T^{D+2}$ term in the low-temperature expansion
of free energy originates in $\delta f_{\text{3loop}}^{(b)}$ of (\ref{3loopFEb}),
but the continuum limit of these  diagrams should be handled with care.
In the direct continuum limit ($a\rightarrow 0$), integral (\ref{3loopFEb}) consists
of two parts. The first one, containing two Bose-distributions, is UV divergent for
all values of $D\geq 3$.
Instead of subtracting this infinite contribution, we may fully exploit the 
lattice regularization to remove the divergent term. If the wave vectors,
whose energies does not appear in Bose factors of (\ref{3loopFEb}), 
are kept within Brilouin zone, we find that first part of $\delta f_{\text{3loop}}^{(b)}$
renormalizes (\ref{FreeEnLatt}), i.e. 
 (\ref{DysonFreeEnT5}): $\delta f_{\text{2loop}} \rightarrow \delta f_{\text{2loop}} [1+Q(D)/S]$.
For the $D$ dimensional simple cubic lattice, the renormalizing factor is 
%-------------------------------------------------------------------------------------------------------------------
\bea
Q(D) = \frac{1}{D} \int_{\bm x} \frac{\cos^2 x_1}{1 - \gamma_D(\bm x)}. \label{IntGama}
\eea
%-------------------------------------------------------------------------------------------------------------------
Here $\bm x$ denotes  dimensionless wave vector within Brilouin zone,
$-\pi \leq x_\alpha \leq \pi, \;\; \alpha = 1,2, \cdots D$.

Now we calculate the finite part of $\delta f_{\text{3loop}}^{(b)}$ in the continuum
limit, the correction to (\ref{DysonFreeEnT5}) of order $T^{(3D+2)/2}$
due to magnon-magnon interactions. Introducing dimensionless wavevectors $\bm x = \bm k /\sqrt{2 m_0 T}, 
\bm y = \bm p /\sqrt{2 m_0 T}$, and  $\bm z = \bm q /\sqrt{2 m_0 T}$, we
obtain the leading order term in $\delta f_{\text{3loop}}^{(b)}$:
%-------------------------------------------------------------------------------------------------------------------
\bea
\hspace*{-0.8cm}
\delta f_{\text{3loop}}^{(b)} (T, h) \approx - \frac{v_0}{\Sigma^2} \left( \frac{\Sigma}{F^2} \right)^{3D/2}
\hspace{-0.1cm} I(D,h)\;
T^{(3D+2)/2} \label{BeyDysH}
\eea
%-------------------------------------------------------------------------------------------------------------------
where
%-------------------------------------------------------------------------------------------------------------------
\bea
I(D,h) = \int_{\bm x, \bm y, \bm z} 
\frac{\Lan n_{\bm x}(h) \Ran  \Lan n_{\bm y}(h) \Ran  \Lan n_{\bm z}(h) \Ran (\bm x \cdot \bm y)^2 }
{\bm z^2 + \bm x \cdot \bm y - \bm z \cdot(\bm x + \bm y)} \label{intD}
\eea
%-------------------------------------------------------------------------------------------------------------------
and
%-------------------------------------------------------------------------------------------------------------------
\bea 
\Lan n_{\bm x}(h) \Ran  = \left[\exp(\bm x^2 + h) - 1  \right]^{-1}, \;\;\; h = \mu H/T.
\eea
%-------------------------------------------------------------------------------------------------------------------
In the absence of external field $H$, (\ref{BeyDysH}) reduces to
%-------------------------------------------------------------------------------------------------------------------
\bea
\delta f_{\text{3loop}}^{(b)} (T) \approx - \frac{v_0}{\Sigma^2} \left( \frac{\Sigma}{F^2} \right)^{3D/2}
\hspace{-0.1cm} I(D)\;
T^{(3D+2)/2} \label{BeyDys}
\eea
%-------------------------------------------------------------------------------------------------------------------
with $I(D) \equiv I(D,h=0)$. The numerical values of integrals $I(D)$ are listed
in the Table \ref{FETab} for $D = 3,4,5$ and 6. 
We note that the value of the coefficient in $T^{11/2}$ term agrees perfectly
with the calculations of \cite{Hofmann3} based on the dimensional regularization.
For more details concerning numerical
evaluation of the integrals $I(D)$, we refer to the Appendix \ref{AppInt}.

The result (\ref{BeyDys}) enable us to calculate corresponding
correction to the spontaneous magnetization induced by magnon-magnon interaction.
Since $\delta \Lan S^z \Ran = - \partial (\delta f_{\text{3loop}})/\partial (\mu H)|_{H = 0}$,
we find
%-------------------------------------------------------------------------------------------------------------------
\bea
\delta \Lan S^z \Ran_{\text{int}}^{\text{3loop}} (T) \approx - 
\frac{v_0}{\Sigma^2} \left( \frac{\Sigma}{F^2} \right)^{3D/2}
\hspace{-0.1cm} J(D)\;
T^{3D/2} \label{BeyDysMagn}
\eea
%-------------------------------------------------------------------------------------------------------------------
where
%-------------------------------------------------------------------------------------------------------------------
\bea
&& J(D) =  \label{intJD} \\
&-&\int_{\bm x, \bm y, \bm z} 
\frac{\Lan n_{\bm x} \Ran  \Lan n_{\bm y} \Ran  \Lan n_{\bm z} \Ran 
[3+\Lan n_{\bm x} \Ran+\Lan n_{\bm y} \Ran+\Lan n_{\bm z} \Ran](\bm x \cdot \bm y)^2 }
{\bm z^2 + \bm x \cdot \bm y - \bm z \cdot(\bm x + \bm y)} \nonumber
\eea
%-------------------------------------------------------------------------------------------------------------------
and $\Lan n_{\bm x} \Ran  = \Lan n_{\bm x} (h=0) \Ran $.
The values of integrals $J(D)$ are also listed in Table \ref{FETab}.
Again, for $D=3$, we find excellent agreement with \cite{Hofmann3}.

%-------------------------------------------------------------------------------------------------------------
\begin{table}
\caption{\label{FETab} Numerical values of integrals $I(D)$ and $J(D)$}
\begin{ruledtabular}
\begin{tabular}{ccc}
$D$  &  $I(D)$                    &       $J(D)$        \\ \hline
$3$  &  $5.3367 \times 10^{-6}$   &     $ -4.401  \times 10^{-5}$       \\ 
$4$  &  $1.2287 \times 10^{-7}$   &     $ -6.565  \times 10^{-7}$       \\ 
$5$  &  $2.08596\times 10^{-9}$   &     $ -1.9410   \times 10^{-9}$       \\
$6$  &  $1.40407\times 10^{-10}$   &    $ -4.9436   \times 10^{-10}$       \\
\end{tabular}
\end{ruledtabular}
\end{table}
%-------------------------------------------------------------------------------------------------------------

At this point we may compare results from three-loop calculations 
for the free energy with those of section \ref{SE2Loop}.
The two leading order terms
in low temperature series for the spontaneous magnetization 
due to magnon-magnon interactions were shown there to carry $D+1$ and $D+2$ powers
of temperature, respectively. Results of three-loop analysis
 for the free energy reveal term with $3D/2$ powers of temperature.
For $D=3$, the $T^{3D/2}$ term indeed represents the first correction
to Dyson's $T^{D+1}$ result. However, already for $D=4$, contribution from
$T^{D+2}$  and   $T^{3 D/2}$ terms  are of the same order. For $D \geq 5$
leading order correction to Dyson-like $T^{D+1}$ term is proportional to
$T^{D+2}$, and is given in (\ref{OneLoopCorr}) and (\ref{Betai}).
Further understanding of relationship between $T^{D+2}$  and   $T^{3 D/2}$ terms
may be reached by directly examining diagrammatic
series for low-temperature expansion of  free energy (see \ref{LoopSeries}).

\section{Symmetry of the effective Lagrangian and Spurious terms in low-temperature expansions}
\label{RotSymLagr}

To gain further insight into the thermodynamics of $D$ dimensional
$\text{O}(3)$ Heisenberg ferromagnet$-$in particular
to see how  various terms in the low-temperature expansion of free energy
are generated by magnon-magnon interactions $-$ we shall now obtain
the  most important  results of Section \ref{3LFE} with the help of
 power counting scheme for magnon fields.
The power counting schemes and structure of diagrams for low-temperature
expansion of the free energy in cases $D=2$ and $D=3$ are given in \cite{Hofmann3,Hofmann4}.
We shall now generalize these results for arbitrary $D\geq 3$.
In this section, the focus will be on the free energy per unit
volume ($f = F/V$), in contrast to the Section \ref{3LFE}, devoted to the calculation
of the free energy per lattice site.

\subsection{Effective Lagrangian and power counting}

Following e.g \cite{DombreRead}, we start from the lattice Hamiltonian (\ref{HamEffLatt0}), 
and obtain continuous
Hamiltonian density up to $|\bm p|^6$, organized in the powers of magnon momenta. We also
include weak external magnetic field $\bm H = H \bm e_z$
%-------------------------------------------------------------------------------------------------------------------
\bea
\H_0 & = & \frac{F^2}{2} \partial_\alpha \bm \pi \cdot \partial_\alpha \bm \pi
- \Sigma \mu H \left( 1- \bm \pi^2/2  \right), \nonumber
%-l_1 \partial_\alpha^2 \bm \pi \cdot \partial_\alpha^2 \bm \pi
%+ l_2 \partial_\alpha^3 \bm \pi \cdot \partial_\alpha^3 \bm \pi \nonumber 
\\ 
\H^{(2)} & = & \frac{F^2}{8} \left[ \bm \pi^2 \bm \pi \cdot \Delta \bm \pi
- \bm \pi^2 \Delta \bm \pi^2  \right]
\equiv \H^{(2)}_{\text{I}} +  \H^{(2)}_{\text{II}}, \nonumber \\
\H^{(4)}  & = & -l_1 \partial_\alpha^2 \bm \pi \cdot \partial_\alpha^2 \bm \pi \label{HContP6} \\
& + & 
 \frac{l_1}{4} \left[ \partial_\alpha^2 \left( \bm \pi^2 \bm \pi \right)
\cdot \partial_\alpha^2 \bm \pi
- \partial_\alpha^2 \bm \pi^2 \partial_\alpha^2 \bm \pi^2 \right], \nonumber \\
\H^{(6)}  & = &  c_1 \partial_\alpha^3 \bm \pi \cdot \partial_\alpha^3 \bm \pi \nonumber \\ 
& + &  \frac{c_1}{4} \left[- \partial_\alpha^3 \left( \bm \pi^2 \bm \pi \right)
\cdot \partial_\alpha^3 \bm \pi
+ \partial_\alpha^3 \bm \pi^2 \partial_\alpha^3 \bm \pi^2 \right]. \nonumber
\eea
%-------------------------------------------------------------------------------------------------------------------
$\H_0$ describes free magnons with rotationally invariant
classical dispersion 
%-------------------------------------------------------------------------------------------------------------------
\bea
E(\bm k)= \frac{F^2}{\Sigma} \bm k^2 + \mu H, \label{ClassRot}
\eea
%-------------------------------------------------------------------------------------------------------------------
and the rest of terms in (\ref{HContP6}) are treated as a perturbation.
While first terms 
in $\H^{(4)}$ and $\H^{(6)}$ modify
dispersion due to the lattice anisotropies, the other contributions include
magnon-magnon interactions.
The formal values for coupling constants 
are $l_1 = F^2 a^2/24, \; c_1 = F^2 a^4/720$.
This Hamiltonian is seen to arise from the Lagrangian
%-------------------------------------------------------------------------------------------------------------------
\bea
\L = \L^{(2)} + \L^{(4)} +\L^{(6)} \label{LContP6}
\eea
%-------------------------------------------------------------------------------------------------------------------
where $\L^{(2)}$ is given in (\ref{EffLagr})
%-------------------------------------------------------------------------------------------------------------------
\bea
\L^{(2)} & = & \Sigma \frac{\partial_t U^1 U^2 - \partial_t U^2 U^1}{1+U^3}
-\frac{F^2}{2} \partial_{\alpha} \bm U \cdot \partial_{\alpha} \bm U \nonumber \\
& + & \Sigma \mu H U^3, \label{EffLagrBold}
\eea
%-------------------------------------------------------------------------------------------------------------------
 and
%-------------------------------------------------------------------------------------------------------------------
\bea
\L^{(4)}  =  l_1 \;  \partial_\alpha^2 \bm U \cdot \partial_\alpha^2 \bm U,
\hspace{0.5cm}
\L^{(6)}  =   c_1 \; \bm U \cdot \partial_\alpha^3 \partial_\alpha^3 \bm U.
 \label{LContP6a}
\eea
%-------------------------------------------------------------------------------------------------------------------
We see that, unlike spin-rotation symmetry, space-rotation symmetry is lost starting from 
$\bm p^4$ terms. 
The crucial point in systematic EFT calculation of the free energy density 
comes from the observation that loop diagrams carry additional powers of momentum
and thus yield terms with increasing powers of temperature.
For the model of current interest, each loop is suppressed
by $D$ powers of momentum, as can bee seen from (\ref{Prop}) 
[See also  \cite{Hofmann2,Hofmann3,Hofmann4}]. This, along with
the fact that $\bm p^2$ counts as $T$, allows for systematic
organization of diagrams contributing free energy density \cite{Gerber}.

Before discussing the low-temperature expansion, we  note that
the single vertex three-lop  diagrams 
%-------------------------------------------------------------------------------------------------------------------
\bea
 \begin{array}{l}
%\vspace{1.1cm} 
\includegraphics[scale=0.8]{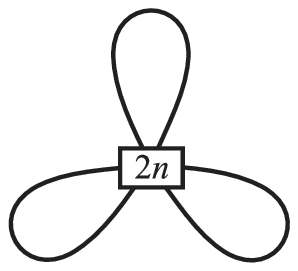},
\end{array}
\eea
%-------------------------------------------------------------------------------------------------------------------
with rectangle denoting vertices from $\H^{(2)}, \H^{(4)}$ or $\H^{(6)}$
(i.e. $n = 1,2,3$),
are completely absent. This is true  for all models containing only terms
quadratic in $\bm U$ and consistent
with symmetries of the Heisenberg model. 
To see how this happens, observe that 
WZW term always produce the six-magnon
vertex with opposite sign than the one 
arising from terms with spatial derivatives only, as can be shown by employing the equation
of motion.
Thus,   for model  (\ref{LContP6}), cancellation of six-magnon vertices is 
exact to all orders in $|\bm p|$.

\subsection{Magnon magnon interactions and low-temperature series for the free energy density}
\label{LoopSeries}

Basically, two types of diagrams contribute to free energy. In the first of them, only
vertices with two magnon fields appear. 
Their general form is
%-------------------------------------------------------------------------------------------------------------------
\bea
 \begin{array}{l}
%\vspace{1.1cm} 
\includegraphics[scale=0.8]{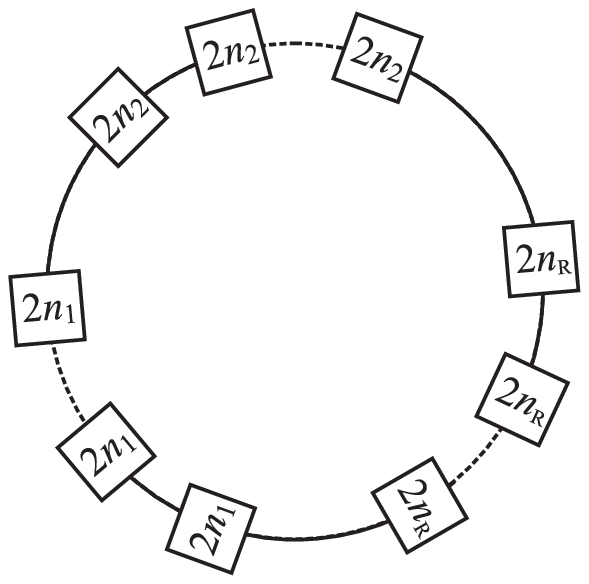} 
\end{array}
\label{FreeLattDiag}
\eea
%-------------------------------------------------------------------------------------------------------------------
where vertex with $2 n_i$ powers of momentum appears $k_i$ times $(i=1,2,\cdots R)$.
As loops are suppressed by $|\bm p|^D$  for Lagrangian (\ref{LContP6}), 
it is easy to see that diagrams of equation (\ref{FreeLattDiag})
contribute to low-temperature expansion of free energy with terms whose power
of $T$ is $D/2+1+\sum_{i=1}^Rk_i(n_i-1)$.
All of them describe the
influence of  lattice anisotropies, i.e. they correspond
to the  $\alpha_i$ coefficients of the low-temperature
expansion of spontaneous magnetization, given in (\ref{DeltaSzFreee}). Since all
two-magnon vertices carry an even power of momentum, contributions
from lattice anisotropies produce the terms with powers 
of temperature equal to $D/2+1, D/2+2$ and so on.

The first diagram that describes magnon-magnon interaction involves a single
 four-magnon vertex of $\H^{(2)}$. By dimensional arguments, it contributes
with a term proportional to $T^{D}$ in low-temperature expansion
of free energy. However, this term vanishes due to the
space-rotation symmetry \cite{Hofmann2,Hofmann3}.
Moreover, we shall now show tat even general class of diagrams,
containing four vertex of $\H^{(2)}$ and two-magnon vertices of $\H^{(2 n)}$
always vanish, i.e. they do not contribute to the free
energy of O(3) HFM. 
To  this end, consider the diagram
%-------------------------------------------------------------------------------------------------------------------
\bea
 \begin{array}{l}
%\vspace{1.1cm} 
\includegraphics[scale=0.8]{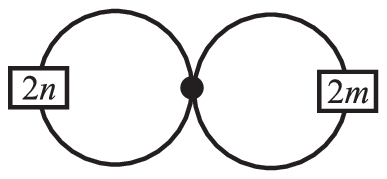} 
\end{array}
\label{ZeroInter}
\eea
%-------------------------------------------------------------------------------------------------------------------
where we have, following standard conventions \cite{Gerber,Hofmann2,Hofmann3}, denoted the vertex
from $\H^{(2)}$ by a dot. The rectangles denote two-magnon vertices
carrying $2n$ and $2m$ powers of $\bm p$, respectively.
First, split the diagram (\ref{ZeroInter}) into two pieces, one
including four-magnon vertex of $\H^{(2)}_{\text I}$ and the other one
with vertex of $\H^{(2)}_{\text{II}}$
(i.e. $ \begin{array}{l}\includegraphics[scale=1.0]{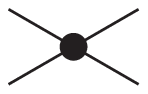}\end{array}
= \begin{array}{l}\includegraphics[scale=1.0]{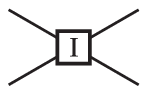}\end{array}
+\begin{array}{l}\includegraphics[scale=1.0]{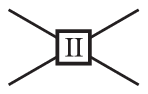}\end{array}$).
$\begin{array}{r}{\includegraphics[scale=1.0]{Fig3_a.eps}}\end{array}$ 
and 
$\begin{array}{r}{\includegraphics[scale=1.0]{Fig3_b.eps}}\end{array}$
bellow refer to $\H^{(2)}_{\text{I}}$ and $\H^{(2)}_{\text{II}}$ of (\ref{HContP6}).
Now
%-------------------------------------------------------------------------------------------------------------------
\bea
 \begin{array}{l}
%\vspace{1.1cm} 
\includegraphics[scale=0.8]{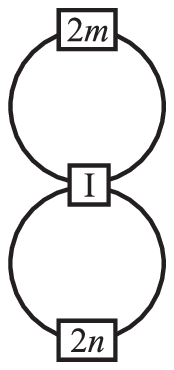}
\end{array} & = & \;\;
\begin{array}{l}
%\vspace{1.1cm} 
\includegraphics[scale=0.8]{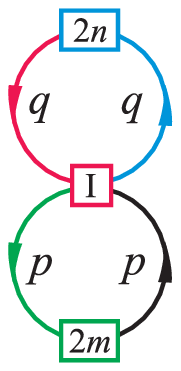}
\end{array}
+ \;\;
\begin{array}{l}
%\vspace{1.1cm} 
\includegraphics[scale=0.8]{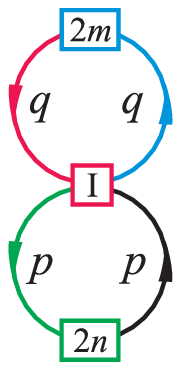}
\end{array} 
 \nonumber \\
& + & \begin{array}{l} 
\includegraphics[scale=0.8]{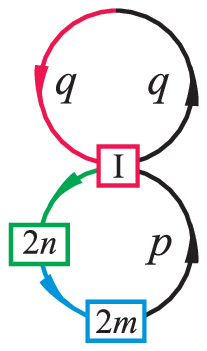}
\end{array}
+    \begin{array}{l}
\includegraphics[scale=0.8]{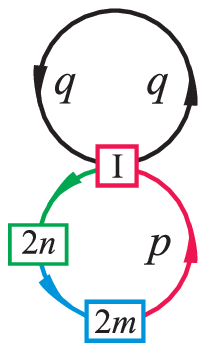}
\end{array},  \label{ZeroInter1}
\eea
%-------------------------------------------------------------------------------------------------------------------
%-------------------------------------------------------------------------------------------------------------------
\bea
 \begin{array}{l}
%\vspace{1.1cm} 
\includegraphics[scale=0.8]{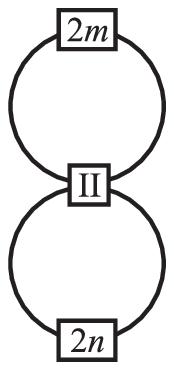}
\end{array} & = & \;\;
\begin{array}{l}
%\vspace{1.1cm} 
\includegraphics[scale=0.8]{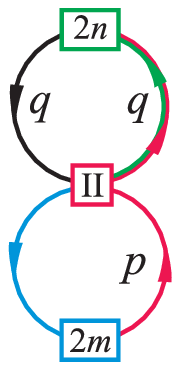}
\end{array}
+ \;\;
\begin{array}{l}
%\vspace{1.1cm} 
\includegraphics[scale=0.8]{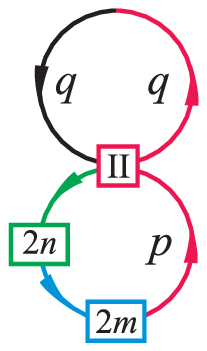}
\end{array} 
\label{ZeroInter2} %\\
\eea
%-------------------------------------------------------------------------------------------------------------------
where first two diagrams in  (\ref{ZeroInter1}) and first one in (\ref{ZeroInter2}) are to be multiplied
 by a factor of 2 and the rest of them by  a factor of four.
(The last two diagrams in (\ref{ZeroInter1}), as well as the second one in (\ref{ZeroInter2})
possess additional symmetry due to permutation of blue and green vertices.) There are four more colored contractions 
in equation (\ref{ZeroInter2}) which we have not displayed, since they  are of the form (\ref{H2OneLoop})
and thus vanish. The diagrams from (\ref{ZeroInter1}) and (\ref{ZeroInter2})
are to be evaluated using diagrammatic  rules from the Section \ref{EffInter},
modified to compensate the replacement of 
lattice magnon dispersion
with (\ref{ClassRot}).
Finally, we see that diagram (\ref{ZeroInter}) is proportional to
%-------------------------------------------------------------------------------------------------------------------
\bea
 \int_{\bm p, \bm q} G(\bm p,\bm q,T)\; \bm p \cdot \bm q
\left[ 2 \text p^{2n} \text p^{2m} + \text p^{2n} \text q^{2m}\right] =0
\label{ZeroInter3}
\eea
%-------------------------------------------------------------------------------------------------------------------
with
%-------------------------------------------------------------------------------------------------------------------
\bea
 G(\bm p,\bm q,T) & = & \frac{\Lan n_{\bm p} \Ran [\Lan n_{\bm p} \Ran +1]
 \Lan n_{\bm q} \Ran [\Lan n_{\bm q} \Ran +1]}{T^2}, \nonumber \\
\text p^{2n} & = & (-)^n \sum_{\alpha=1}^D p_{\alpha}^{2n}. \nonumber
\eea
%-------------------------------------------------------------------------------------------------------------------
A special cases of vanishing diagram (\ref{ZeroInter}) are the ones with
single  two-magnon vertex or with four-magnon vertex only.

\vspace*{0.5cm}

In a similar manner, one can also demonstrate the vanishing of 
three-loop diagram
%-------------------------------------------------------------------------------------------------------------------
\bea
 \begin{array}{l}
%\vspace{1.1cm} 
\includegraphics[scale=0.8]{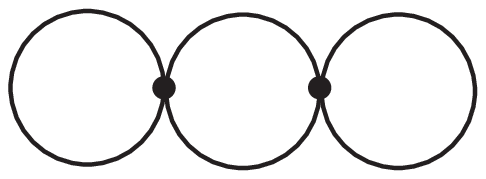} 
\end{array}
\label{3LoopZeroGraph}
\eea
%-------------------------------------------------------------------------------------------------------------------
(see also \cite{Hofmann3} and discussion just before (\ref{3LoopZero})).

Thus, a list  of different diagrams arising from magnon-magnon interactions
 that actually contribute to
the free energy of O(3) HFM is quite limited. The first term in the
low-temperature expansion of the free energy density, carrying $D+2$ powers
of temperature (the Dyson term), is generated by the two-loop diagram
%-------------------------------------------------------------------------------------------------------------------
\bea
 \begin{array}{l}
%\vspace{1.1cm} 
\includegraphics[scale=0.8]{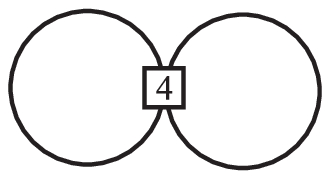} 
\end{array}
\label{TempD+2}.
\eea
%-------------------------------------------------------------------------------------------------------------------
Its value is 
%-------------------------------------------------------------------------------------------------------------------
\bea
\delta f_{D+2} & = & - \frac{3 l_1 D \pi^D}{2 \Sigma^2} \left(\frac{1}{2 \pi} \right)^{2D}
 \left( \frac{\Sigma }{F^2} \right)^{D+2}  T^{D+2} \nonumber \\
 & \times & \left[ \sum_{n = 1}^{\infty} \frac{\e^{-\mu H n /T}}{n^{(D+2)/2}} \right]^2.
 \label{ContFreeEnT5}
\eea
%-------------------------------------------------------------------------------------------------------------------
Out of the two-loop graphs, now comes the $T^{D+3}$ term. It is given by
%-------------------------------------------------------------------------------------------------------------------
\bea
 \begin{array}{l}
%\vspace{1.1cm} 
\includegraphics[scale=0.8]{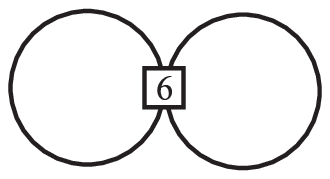} 
\end{array}
+
 \begin{array}{l}
%\vspace{1.1cm} 
\includegraphics[scale=0.8]{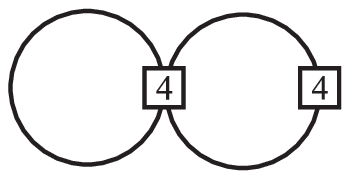} 
\end{array}
\label{TempD+3},
\eea
%-------------------------------------------------------------------------------------------------------------------
so that
%-------------------------------------------------------------------------------------------------------------------
\bea
\delta f_{D+3} & = & - \frac{3 D \pi^D}{8 \Sigma^2} \left(\frac{1}{2 \pi} \right)^{2D}
 \left( \frac{\Sigma }{F^2} \right)^{D+3}  T^{D+3}
 \nonumber \\
 & \times & \left[\frac{12 l_1^2[D+4]}{F^2} -30 c_1 \right] \nonumber \\
 & \times &   \sum_{n,m = 1}^{\infty} \frac{\e^{-\mu H [n+m] /T}}{n^{(D+2)/2}\; m^{(D+4)/2}} .
 \label{ContFreeEnT6}
\eea
%-------------------------------------------------------------------------------------------------------------------
Next enter three-loop diagrams. The lowest non-zero diagram is
%-------------------------------------------------------------------------------------------------------------------
\bea
 \begin{array}{l}
%\vspace{1.1cm} 
\includegraphics[scale=0.8]{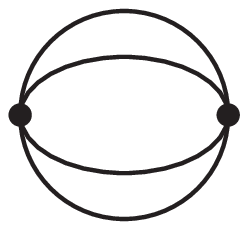} 
\end{array}
\label{Temp3D/2+1}.
\eea
%-------------------------------------------------------------------------------------------------------------------
This diagram contains an infinite contribution (See the Section \ref{3LFE}).
Its finite part is given by
%-------------------------------------------------------------------------------------------------------------------
\bea
\delta f_{3D/2+1} & = &  - \frac{1}{\Sigma^2} \left( \frac{\Sigma}{F^2} \right)^{3D/2}
\hspace{-0.1cm} I(D,h)\;
T^{(3D+2)/2},
 \label{ContFreeEnT11-2}
\eea
%-------------------------------------------------------------------------------------------------------------------
where numbers $I(D,h)$ are defined in (\ref{intD}).
If the external magnetic field is switched off, 
equations (\ref{ContFreeEnT5}),
(\ref{ContFreeEnT6}) and (\ref{ContFreeEnT11-2})
give three
 lowest contributions to free energy density due to magnon-magnon interactions for
$D=3,4$ and 5-dimensional ferromagnets. 
It was already mentioned, in the context of spontaneous magnetization,
 that $T^{3 D/2+1}$ term of (\ref{ContFreeEnT11-2})
represents the leading correction to the Dyson's $T^{D+2}$ term only
for $D=3$. This  is understood quite naturally
in the language of EFT: The $T^{3 D/2+1}$-term comes from 
three-loop graph (\ref{Temp3D/2+1}), and with increasing $D$ it is being pushed
up in the temperature expansion compared to two-loop graphs
(\ref{TempD+3}).
At $D=6$, one also
needs to include four-magnon vertices of order $|\bm p|^8$ 
to calculate $T^{D+4}$ term from two-loop graphs, which is 
of the same order as $T^{3D/2+1}$ term of (\ref{Temp3D/2+1}).

\subsection{Spurious terms}
\label{ST}

Now we examine in detail what is, in the  section \ref{EFTRPA},
 shown to be the effective field theory  for TRPA.
The  Hamiltonian density obtained from (\ref{HamEffLattRPA}), up to the terms
of order $|\bm p|^6$ reads
%-------------------------------------------------------------------------------------------------------------------
\bea
\H^{(2)}_{\text{RPA}} & = & \frac{F^2}{8} 
 \bm \pi^2 \Delta \bm \pi^2, \nonumber \\
\H^{(4)}_{\text{RPA}}  & = & -l_1 \partial_\alpha^2 \bm \pi \cdot \partial_\alpha^2 \bm \pi
+
 \frac{l_1}{4} 
\partial_\alpha^2 \bm \pi^2 \partial_\alpha^2 \bm \pi^2, \label{HContP6RPA}  \\
\H^{(6)}_{\text{RPA}}  & = &  c_1 \partial_\alpha^3 \bm \pi \cdot \partial_\alpha^3 \bm \pi +
  \frac{c_1}{4} 
 \partial_\alpha^3 \bm \pi^2 \partial_\alpha^3 \bm \pi^2. \nonumber
\eea
%-------------------------------------------------------------------------------------------------------------------
$\H_0$ and the coefficients $l_1$ and $c_1$  are the same as ones in (\ref{HContP6}).
The corresponding Lagrangian up to $|\bm p|^6$ consists of three parts,
$\L_{\text{RPA}} = \L^{(2)}_{\text{RPA}} + \L^{(4)}_{\text{RPA}} +\L^{(6)}_{\text{RPA}}$, where
each term contains contributions that explicitly break internal symmetry of the
Heisenberg Hamiltonian (\ref{HFM1})
%-------------------------------------------------------------------------------------------------------------------
\bea
\L_{\text{RPA}}^{(2)}   & = & \frac \Sigma 2 \left( \partial_t U^1 U^2 - \partial_t U^2 U^1\right)
-\frac{F^2}{2} \partial_{\alpha} \bm U \cdot \partial_{\alpha} \bm U \nonumber \\
& + & \Sigma \mu H U^3 - \frac{F^2}{4}  \bm \pi^2 \Delta \bm \pi^2, \label{EffLagrBoldRPA} \\
\L^{(4)}_{\text{RPA}} & = & l_1 \partial_\alpha^2 \bm U \cdot \partial_\alpha^2 \bm U
- \frac{l_1}{2} \bm \pi^2 \partial_\alpha^2 \partial_\alpha^2 \bm \pi^2, \nonumber \\
\L^{(6)}_{\text{RPA}}  & = & l_1 \partial_\alpha^3 \bm U \cdot \partial_\alpha^3 \bm U
- \frac{l_1}{2} \bm \pi^2 \partial_\alpha^3 \partial_\alpha^3 \bm \pi^2 \nonumber.
\eea
%-------------------------------------------------------------------------------------------------------------------

It is easily seen from (\ref{HContP6RPA})  that
TRPA contains the same one-loop diagrams (\ref{FreeLattDiag}) 
as  correct effective theory for HFM defined by  (\ref{HContP6}).
The most important difference in low temperature expansion concerns the 
two-loop diagram of four-magnon vertex $\H^{(2)}_{\text{RPA}}$.
It is now given by
%-------------------------------------------------------------------------------------------------------------------
\bea
 \begin{array}{l}
%\vspace{1.1cm} 
\includegraphics[scale=0.8]{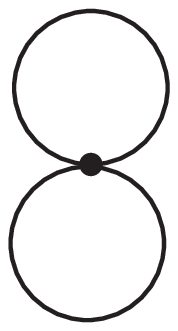} 
\end{array}
&=&
 \begin{array}{l}
%\vspace{1.1cm} 
\includegraphics[scale=0.8]{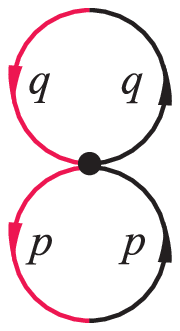} 
\end{array}
+
 \begin{array}{l}
%\vspace{1.1cm} 
\includegraphics[scale=0.8]{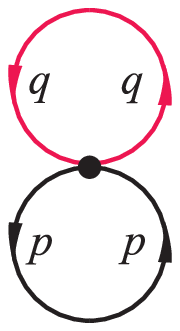} 
\end{array}  \label{RPAD+1Graph} 
\eea
%-------------------------------------------------------------------------------------------------------------------
so that
%-------------------------------------------------------------------------------------------------------------------
\bea
\delta f^{\text{RPA}}_{D+1}
&=& - \frac{D \pi^D}{2 \Sigma} \left( \frac{1}{2 \pi} \right)^{2D}
\left( \frac{\Sigma}{F^2} \right)^{2D} T^{D+1} \nonumber \\
&\times& \sum_{n,m = 1}^{\infty} \frac{\e^{-\mu H [n+m] /T}}{n^{(D+2)/2}\; m^{D/2}}  
\label{RPAD+1}.
\eea
%-------------------------------------------------------------------------------------------------------------------
The dot in (\ref{RPAD+1Graph}) represents four-magnon vertex of
$\H^{(2)}_{\text{RPA}}$ from (\ref{HContP6RPA}).
Clearly, (\ref{RPAD+1}) is responsible   for the spurious $T^{D+1}$  term
(See the equations (\ref{DeltaSzRPA2}) and (\ref{DeltaSzRPA3})).
In this manner, we have reached the same conclusion as in lattice regularized theory:
the main reason for spurious $T^{D+1}$ term in low-temperature series for free energy
 is the explicit violation of internal
symmetry of HFM by discarding magnon-magnon interactions
from the WZW term. The effective action for TRPA Lagrangian is no longer O(3) invariant.
Even though it was shown that space-rotation symmetry of leading order
Lagrangian guaranties absence of $T^{D+1}$ term (See \cite{Hofmann2,Hofmann3}
for $D=3$),
we see that  it only does so if
the full spin-rotational symmetry of effective action is preserved (order by order, in the
perturbative expansion).

The importance of WZW term in effective Lagrangian for ferromagnet
was stressed in \cite{PRD,WenZee} where it 
was shown that it is crucial for obtaining correct qualitative
magnon spectrum in linear approximation.
The present paper puts conclusions of \cite{PRD,WenZee}
a step forward in the sense that it is demonstrated how the thermodynamical
properties of a ferromagnet  depend 
on the magnon-magnon interactions generated by  WZW term.

\section{Sumarry}

The method of effective Lagrangians is 
a powerful
tool for analyzing  the low-energy (low temperature) domain of 
models with strongly interacting constituents,
provided 
spontaneous symmetry breaking occurs.
Employed first for the description of low-energy sector of QCD,
it soon  found its way to the
strongly correlated systems of condensed  matter.
A notable example
for the latter class of 
systems is the isotropic Heisenberg ferromagnet,
exhibiting $\text O(3)\longrightarrow \text O(2)$ spontaneous breakdown of 
internal spin-rotation symmetry.
Due to a symmetry breakdown,
massless particles (Goldstone bosons)
appear in the spectrum. In the case of Heisenberg magnets,
Goldstone bosons are  
called magnons and for O(3)
HFM, there are two broken generators and a single magnon. 
The EFT line of reasoning then constitutes in writting the 
most general Lagrangian (i.e. action) consistent with
internal symmetries of the Heisenberg Hamiltonian,
organized in the powers of  momenta of Goldstone fields. 
By employing 
appropriate Feynman rules for time-ordered Green's functions
of the effective Lagrangian, one can calculate free energy
and other physical quantities of interest.

EFT  is usually formulated in the
functional integral framework using dimensional 
regularization.
Lattice regularization, however,  enables the effective Hamiltonian to
inherit the full
discrete symmetry of the original Heisenberg ferromagnet 
which modifies the free  magnon dispersion from
$\bm k^2/(2m_0)$ to $\widehat{\bm k}^2/(2m_0)$.
In the spirit of chiral perturbation theory, where original
quark and gluons of QCD are replaced with Goldstone bosons of 
spontaneously broken chiral symmetry, no reference  is 
made to the original degrees of freedom, i. e. to the localized spins.
The theory includes  bosonic (magnon) fields from the  beginning.
Because of that, calculations in lattice EFT are free
of certain types of  approximations unavoidable in 
approaches relaying on equations of motion for spin operators
or on their bosonic/fermionic representations.
A vital part of our analysis is careful
inclusion of the magnon-magnon interactions arising from
the WZW term in the effective Lagrangian.

Calculations in EFT  reduce to evaluation of various
Feynman diagrams involving propagators of Goldstone fields.
As Goldstone fields are derivatively coupled, the lattice regularization
necessary induces diagrams containing at least one lattice
Laplacian acting on propagators. The number of lattice Laplacians increases with
the order of diagram, i.e. with number of loops. 
The structure of diagrams becomes even more involved if 
interaction part of Hamiltonian consists 
of several terms  different in derivative structure
since more than one lattice Laplacian may act
on given propagator, as in the present paper.
To get around these complications as much as possible, we have devised
a variant of Feynman diagrams suitable for scalar fields 
with derivative couplings. They are applicable for lattice 
as well for continuum field theories. Corresponding Feynman
rules are based on colored lines and vertices and are
exposed in Section \ref{Diagrammar}. All calculations in the present
paper  rely on this version of Feynman diagrams.

Once the  effective  Hamiltonian of lattice magnon fields is
found, it can 
be used to study magnon-magnon interactions and their influence
on magnon self energy, spontaneous magnetization and free energy
of $D\geq 3$ HMF.
We have shown in Section \ref{SE2Loop} that the Dyson's correction to the
spontaneous magnetization can be obtained from
one-loop correction to the magnon propagator, 
bypassing discussion on  the free energy.
The  approach simplifies calculations
and opens direct link to the standard theories of spin waves in the
Heisenberg ferromagnet.
In particular, using EFT, we have established connection
between theories on nonlinear spin waves, Tyablikov GF
method and importance of magnon-magnon interactions
originating in the WZW term. 
It is also shown in 
Section \ref{SE2Loop} that, similarly to pions in 
Lorentz-invariant models, ferromagnetic magnons remain "massless"
at two-loop. 

Further analysis of three-loop corrections to the free energy of HFM,
presented in Sections \ref{3LFE} and \ref{RotSymLagr},
reveals how magnon-magnon interactions manifest themselves
in the low-temperature series for free energy.
The two-loop correction yields  terms proportional
to $T^{D+2}$, $T^{D+3}$ and so on, while the three-loop
corrections start with term proportional to $3D/2+1$
powers of temperature. Discussion from Section \ref{RotSymLagr} shows how
interplay between the number of loops and the spatial dimensionality
of the lattice dictates structure of this low-temperature series. 
We also present numerical values for coefficients 
of $T^{3D/2+1}$ term in case of 3,4,5 and 6-dimensional lattice along
with general expression valid for all $D\geq 3$.
While results for $D=3$ are found to be in 
excellent agreement with recent literature, those
 for $D=4,5$ and 6 are completely new.
The use of Hamiltonian formalism leads to  certain simplifications in diagrammatic series for
the free energy, as well as for the magnon self-energy,
since six-magnon vertices precisely cancel
to all orders in $|\bm p|$. Also, using colored diagrams,
we have shown that certain two and three-loop diagrams
for free energy vanish.
At the end of Section \ref{RotSymLagr}
we show that spurious
$T^D$ term in the low-temperature expansion of 
spontaneous magnetization characterizes theories inconsistent with
 internal O(3) symmetry of $D$ dimensional HFM.
For example, explicit breakdown of O(3) symmetry in Tyablikov RPA
is caused by neglection of magnon-magnon interactions
of the form $\bm \pi^2 \bm \pi \cdot \nabla^2 \bm \pi$
 generated solely by the WZW term, thus making
TRPA a
theory of pure bosonic lattice Schr\"{o}dinger field. 
This becomes especially clear when TRPA effective
theory is put in the form given in the Appendix \ref{AppRPA}.

The Heisenberg model and the spin waves on hypercubic lattice have been studied
in the past. These  papers, however, 
focus on general properties of magnetic systems on hypercubic lattices,
such as existence of the Landau-Lifshitz equation \cite{Moser}. Also, their analysis
relies on MF approximations \cite{AuerbachArovas},
 coupled-cluster \cite{Bishop}
and $1/Z_1$ \cite{Fishman} expansion or RG methods \cite{CHN,Kopietz}.
In contrast, we investigate the low-energy magnon-magnon interactions in detail.
Explicit expressions describing
their influence on the magnon self energy,  the ferromagnet free energy
and the spontaneous magnetization are presented in the paper.
We also discuss  subtleties  concerning the
influence of spatial dimensionality  of the lattice
and the number of loops  entering  the low-temperature expansion
of ferromagnet free energy, which can not  be found in earlier works.

To conclude, we may say that the lattice regularization
offers new perspectives on EFT approach to Heisenberg 
ferromagnet. By keeping full magnon dispersion it extends
EFT range of validity at not too low temperatures.
It also makes comparison between EFT results and
those of nonlinear spin waves/TRPA direct and precise.
Finally, although the lattice regularization for effective Lagrangians of more complex
models may not be as straightforward
as for HFM,
there seems to be no principal obstacle in extending  this method
to other systems.

\section*{Acknowledgement}

This work was supported by the Serbian Ministry of Education and Science,
 Project OI 171009. The authors acknowledge the use of the Computer Cluster of the Center 
for Meteorology and Environmental Predictions  of the Department of 
Physics, Faculty of Sciences, University of Novi Sad, Novi Sad, Serbia.

\appendix
\section{Integrals  $I(D)$ and $J(D)$}\label{AppInt}

In this Appendix we give some details concerning
evaluation of integrals $I(D)$ and $J(D)$ that appear as 
numerical prefactors of  $T^{(3 D+2)/2}$ term in the low-temperature
series for free energy and $T^{3D/2}$ term of low-temperature
expansion of spontaneous magnetization, respectively.

The integrals in question are defined in (\ref{intD})
for $h=0$  and in (\ref{intJD}).
The $3D$-fold integration may be reduced to 4-fold one
using $D-$dimensional spherical coordinates %[NEKA REFERENCA]
%-------------------------------------------------------------------------------------------------------------------
\bea
x_1 & = & x \sin \theta_{D-1}  \sin \theta_{D-2} \cdots \sin \theta_2  \sin \theta_1, \nonumber \\
x_2 & = & x \sin \theta_{D-1}  \sin \theta_{D-2} \cdots \sin \theta_2  \cos \theta_1, \nonumber \\
x_3 & = & x \sin \theta_{D-1}  \sin \theta_{D-2} \cdots \sin \theta_3  \cos \theta_2, \nonumber \\
x_4 & = & x \sin \theta_{D-1}  \sin \theta_{D-2} \cdots \sin \theta_4  \cos \theta_3, \nonumber \\
& \vdots & \nonumber \\
x_D & = & x \cos \theta_{D-1}
\eea
%-------------------------------------------------------------------------------------------------------------------
with $x = |\bm x|$,   $0 \leq \theta_1 \leq 2 \pi$ and all other angles
ranging from $0$ to $\pi$. 
By appropriate change of variables, the  $\theta_{D-1}$ integral  of $\bm z$ vector
may be reduced to the integral representation of hypergeometric function
\cite{GiR}
%-------------------------------------------------------------------------------------------------------------------
\bea
_2 F_1 \left( 1, \frac{D-1}{2}; D-1; \frac{2 z |\bm x + \bm y|}
{\bm z^2 + \bm x \cdot \bm y + z |\bm x + \bm y|} \right).
\eea
%-------------------------------------------------------------------------------------------------------------------
Out of remaining $3D-4$ angles, $3D-5$  may be directly integrated, thus 
leaving four dimensional integrals. 
Explicit form of $I(D)$, 
 for $D = 3,4,5$ and $D=6$, are listed below
%-------------------------------------------------------------------------------------------------------------------
\begin{widetext}
\bea
I(3) & = & 2 \left( \frac{1}{2 \pi}  \right)^6 \int_0^\infty \d x  \int_0^\infty \d y     \int_0^\infty \d z
\; x^4 y^4 z \Lan n_x \Ran   \Lan n_y \Ran   \Lan n_z \Ran 
\int_0^\pi \d \theta \sin \theta \frac{\cos^2 \theta}{\lambda(x,y,\theta)} 
\ln \left|\frac{A(x,y,z,\theta)}{B(x,y,z,\theta)} \right|, \label{I3} 
\eea
\bea
I(4) & = & \frac 12 \left( \frac{1}{2 \pi}  \right)^7 \int_0^\infty \d x  \int_0^\infty \d y     \int_0^\infty \d z
\; x^5 y^5 z \Lan n_x \Ran   \Lan n_y \Ran   \Lan n_z \Ran  \nonumber \\
& \times & \int_0^\pi \d \theta \sin^2 \theta \frac{ B(x,y,z,\theta)  \cos^2 \theta}{\lambda^2(x,y,\theta)}
\left[1-\exp\left(\frac 12 \ln \left| \frac{A(x,y,z,\theta)}{B(x,y,z,\theta)} \right| \right)   \right]^2 , \label{I4} 
\eea
\bea
I(5) & = & \frac{1}{6} \left( \frac{1}{2 \pi}  \right)^9 \int_0^\infty \d x  \int_0^\infty \d y     \int_0^\infty \d z
\; x^6 y^6 z \Lan n_x \Ran   \Lan n_y \Ran   \Lan n_z \Ran  \nonumber \\
& \times & \int_0^\pi \d \theta \sin^3 \theta \frac{\cos^2 \theta}{\lambda^3(x,y,\theta)}
\left[2 z \lambda(x,y,\theta) \left(z^2 +x y \cos \theta\right) -
 A(x,y,z,\theta) B(x,y,z,\theta) \ln \left| \frac{A(x,y,z,\theta)}{B(x,y,z,\theta)} \right|  \right] , \label{I5} 
\eea
\bea
I(6) & = & \frac{1}{36} \left( \frac{1}{2 \pi}  \right)^{10} \int_0^\infty \d x  \int_0^\infty \d y     \int_0^\infty \d z
\; x^7 y^7 z \Lan n_x \Ran   \Lan n_y \Ran   \Lan n_z \Ran  \;
\int_0^\pi \d \theta \sin^4 \theta \frac{B^3(x,y,z,\theta) \cos^2 \theta}{\lambda^4(x,y,\theta)}  \nonumber \\ 
& \times & \left[ \exp\left(\frac 32 \ln \left| \frac{A(x,y,z,\theta)}{B(x,y,z,\theta)}  \right|\right)-1     
+ \frac{z \lambda(x,y,\theta)}{B(x,y,z,\theta)}
  \left\{ 3 - \frac 32 \frac{z \lambda(x,y,\theta)}{B(x,y,z,\theta)} -
  \frac 12 \left( \frac{z \lambda(x,y,\theta)}{B(x,y,z,\theta)}  \right)^2    \right\}   \right],  \label{I6}
\eea
\end{widetext}
%-------------------------------------------------------------------------------------------------------------------
where we have introduced shorthand notations common to all integrands
%\newpage
%-------------------------------------------------------------------------------------------------------------------
\bea
\Lan n_x \Ran & = & \frac{1}{\e^{x^2}-1}, \;\;\; \text{etc} \nonumber
\eea
\bea
\lambda(x,y,\theta) & = & \sqrt{x^2 + y^2 + 2 x y \cos\theta }  \nonumber 
\eea
\bea
A(x,y,z,\theta) & = & z^2 + x y \cos \theta - z \lambda(x,y,\theta) \nonumber 
\eea
\bea
B(x,y,z,\theta) & = & z^2 + x y \cos \theta + z \lambda(x,y,\theta).
\eea
%-------------------------------------------------------------------------------------------------------------------
Remaining integrals are rapidly convergent and may be evaluated, e.g.,  by  Gaussian quadrature
(See \cite{NRF} and references therein). The numerical
values for $I(3)$, $I(4)$, $I(5)$ and $I(6)$ are given in the main text
(Table \ref{FETab}).

Integrals $J(D)$, appearing as numerical coefficients of  low-temperature
expansion of spontaneous magnetization are obtained by replacing
$\Lan n_x \Ran   \Lan n_y \Ran   \Lan n_z \Ran$ with
%-------------------------------------------------------------------------------------------------------------------
\bea 
- \Lan n_x \Ran     \Lan n_y \Ran   \Lan n_z \Ran  
[3+\Lan n_x \Ran + \Lan n_y \Ran + \Lan n_z \Ran]
\eea
%-------------------------------------------------------------------------------------------------------------------
in (\ref{I3})-(\ref{I6}). Their numerical values are listed in
the Table \ref{FETab}.

\section{TRPA as the two-body Schr\"{o}dinger theory} \label{AppRPA}

In this Appendix, we reformulate
results of the Subsection \ref{EFTRPA} in the standard 
language of diagrammatic perturbation theory
for nonrelativistic bosons, i.e. without
derivative coupling of magnon fields and Feynman rules of 
\ref{Diagrammar}.
 By simple manipulation we cast  $\widetilde{H}_{\text{int}} = -H_{\text{int}}^{(\text{II})}$
 in the usual form of a two-body  interaction
for the Schr\"odinger field \cite{Wen}
%------------------------------------------------------------------------------------------------------------------------
\bea
\widetilde H_{\text{int}} = \frac{v_0^2}{2} \sum_{\bm x,\; \bm y}
\psi^\dagger(\bm x) \psi^\dagger(\bm y) V(\bm x - \bm y) \psi(\bm y) \psi(\bm x),
\label{TwoBodyRPA}
\eea
%------------------------------------------------------------------------------------------------------------------------
where 
%------------------------------------------------------------------------------------------------------------------------
\bea
\hspace*{-0.1cm}
 V(\bm x - \bm y) = \hspace{-0.1cm} 
 \frac{2 D [F/\Sigma]^2}{Z_1 |\bm \lambda|^2 v_0 } \hspace*{-0.1cm}  \sum_{\bm \lambda}
\hspace*{-0.1cm}  \Big{[}   \Delta(\bm y = \bm x + \bm \lambda) -  \Delta(\bm y = \bm x) \Big{]}. \nonumber
\eea
%------------------------------------------------------------------------------------------------------------------------
The one-loop magnon self energy for this model can
be written as
%------------------------------------------------------------------------------------------------------------------------
\bea
\hspace*{-0.3cm}\widetilde{\Sigma}(\bm k,\omega) &=& \hspace{-0.2cm} \begin{array}{l}
\vspace{1.4cm} \includegraphics[scale=0.7]{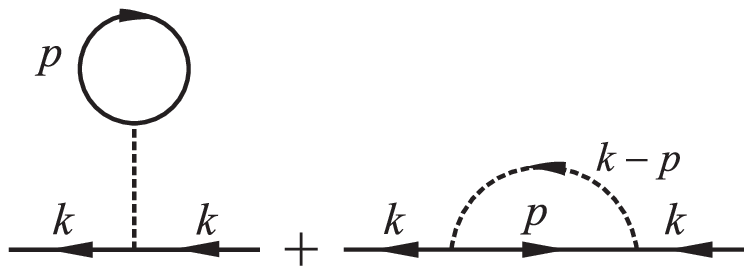}
\end{array}.  %\begin{array}{l}
\label{Diag11RPA}
\eea
%------------------------------------------------------------------------------------------------------------------------
\vspace*{-1.1cm}

\noindent The dashed line here represents the (minus of) Fourier transform of the
two-body potential, $V(\bm k) = - 2  D F^2/(|\bm \lambda|^2 \Sigma^2) [1-\gamma(\bm k)]$.
All results of the Subsection \ref{EFTRPA} could  be
deduced equally well starting from (\ref{TwoBodyRPA})-(\ref{Diag11RPA}).

\end{document}